\documentclass{aa}

\usepackage{graphicx}
\usepackage{txfonts}
\usepackage{arydshln}
\usepackage[colorlinks=true,citecolor=blue,urlcolor=blue]{hyperref}
\usepackage{url}
\usepackage{multirow}

\newcommand{\ubar}[1]{\text{\b{$#1$}}}

\begin{document}

\title{Clustering properties of TGSS radio sources}
\titlerunning{Clustering of TGSS radio sources}

\author{Arianna Dolfi\inst{\ref{Swinburne},\ref{Rome}}
\and Enzo Branchini\inst{\ref{Rome},\ref{INFN},\ref{INAF-R}}
\and Maciej Bilicki\inst{\ref{Leiden},\ref{CFT}} 
\and \\ Andr\'{e}s Balaguera-Antol\'{\i}nez\inst{\ref{IAC},\ref{Laguna}}
\and Isabella Prandoni\inst{\ref{INAF-B}}
\and Rishikesh Pandit\inst{\ref{Rome}}
}

\institute{
Centre for Astrophysics and Supercomputing, Swinburne University of Technology, Hawthorn VIC 3122, Australia \label{Swinburne}
\and
Dipartimento di Matematica e Fisica, Universit\'{a} degli Studi Roma Tre, Via della Vasca Navale, 84, 00146 Roma, Italy \label{Rome}
\and
INFN - Sezione di Roma Tre, via della Vasca Navale 84, I-00146 Roma, Italy \label{INFN}
\and
INAF - Osservatorio Astronomico di Roma, via Frascati 33, I-00040 Monte Porzio Catone (RM), Italy \label{INAF-R}
\and
Leiden Observatory, Leiden University, PO Box 9513, NL-2300RA Leiden, the Netherlands \label{Leiden}
\and
Center for Theoretical Physics, Polish Academy of Sciences, al. Lotnik\'{o}w 32/46, 02-668, Warsaw, Poland \label{CFT}
\and
Instituto de Astrof\'{\i}sica de Canarias, s/n, E-38205 La Laguna, Tenerife, Spain \label{IAC}
\and
Departamento de Astrof\'{\i}sica, Universidad de La Laguna, E-38206 La Laguna, Tenerife, Spain \label{Laguna}
\and
INAF-Instituto di Radioastronomia, Via P. Gobetti 101, I-40129 Bologna, Italy \label{INAF-B}
}

\authorrunning{A. Dolfi et al.}

\offprints{A.~Dolfi, \email{adolfi@swin.edu.au}}

\date{Received: 25 September 2018. Accepted: 17 January 2019}

\abstract{
We investigate the clustering properties of radio sources in the Alternative Data Release 1 of the  TIFR GMRT Sky Survey (TGSS),
focusing on large angular scales, where previous analyses have detected a large clustering signal.
After appropriate data selection, the TGSS sample we use contains $\sim 110,000$ sources selected at 150 MHz over $\sim 70 \%$ of the sky. The survey footprint is largely superimposed 
on that of the NRAO VLA Sky Survey (NVSS) with the majority of TGSS sources having a counterpart in the NVSS sample.
These characteristics make TGSS suitable for large-scale clustering analyses and facilitate the comparison 
with the results of previous studies.
In this analysis we focus on the angular power spectrum, although the angular correlation 
function is also computed to quantify the contribution of multiple-component radio sources.
We find that on large angular scales, corresponding to multipoles $2 \leq \ell \leq 30$, the amplitude of the 
TGSS angular power spectrum is significantly larger than that of the NVSS.
We do not identify any observational systematic effects that may explain this mismatch.
We have produced a number of physically motivated models for the TGSS angular power spectrum
and found that all of them fail to match observations, even when taking into account 
observational and theoretical uncertainties. The same models provide a good fit to the 
angular spectrum of the NVSS sources.
These results confirm the anomalous nature of the TGSS large-scale power, which has no 
obvious physical origin and  seems to indicate that unknown systematic errors are present in the TGSS dataset.}


\maketitle 



\section{Introduction}
\label{sec:intro}

Imaging of the sky at radio frequencies is one of the possible approaches to studying the nature and cosmological evolution of radio sources and their relation to the underlying
large-scale structure (LSS) of the universe. Outside the plane of our Galaxy, most of the sources detected at centimeter and meter wavelengths are extragalactic and often at very high redshifts. This is related to the emission mechanisms at such frequencies, which are nonthermal and occur in specific environments where electrons are accelerated to relativistic velocities and produce synchrotron radiation. The observed extragalactic radio sources are therefore hosts of powerful engines such as active galactic nuclei or sites of intensive star formation, and can be detected from very large cosmological distances. 
This, together with the fact that they are unaffected by dust extinction, makes extragalactic radio sources very useful to probe large cosmological volumes.

Amongst existing radio catalogs, a few wide-angle, sub-arc-minute-resolution catalogs cover areas up to thousands of square degrees that can be suitably used for LSS studies.
Some notable examples include the Green Bank survey at 4.85 GHz  \citep[87GB,][]{87GB}, the Parkes-MIT-NRAO survey also at 4.85 GHz \citep [PMN,][]{PMN}, the Faint Images of the Radio Sky at Twenty centimeters \citep[FIRST,][]{FIRST}, the Westerbork Northern Sky Survey at 325 MHz \citep[WENSS,][]{WENSS},  the NRAO VLA Sky Survey at 1.4 GHz \citep[NVSS,][]{NVSS}, or the Sydney University Molonglo Sky Survey at 843 MHz \citep[SUMSS,][]{SUMSS}. More recently large swaths of sky have been mapped  by the Giant Metrewave Radio Telescope \citep[GMRT,][]{GMRT}, the
Low-Frequency Array \citep[LOFAR,][]{LOFAR}, and the Murchison Widefield Array  \cite[MWA,][]{MWA}.
In the near future, such efforts are expected to accelerate the growth of wide-angle radio datasets by orders of magnitude thanks to forthcoming surveys such as the VLA Sky Survey\footnote{\url{https://science.nrao.edu/science/surveys/vlass}} \citep[VLASS,][]{VLASS} or those that will be undertaken by the Square Kilometre Array \citep[SKA,][]{SKA,SKAsurveys} and its precursors \citep[see e.g.,][]{EMU}.

Studying LSS with radio imaging raises some specific challenges. The nonthermal character of radio emission means that the observed intensity of radio sources is only very weakly related to their distances, unlike in the optical where the bulk of the flux is black-body-like and hence readily provides information about luminosity distance.
Another issue is related to the often complicated morphology of the radio sources. While usually point-like or at least concentrated to a small ellipse at short wavelengths, in the radio domain galaxies often present double or multiple structure with very extended lobes 
that generates clustering signal on small scales, with 
optical/IR counterparts that are difficult to identify. Furthermore, radio galaxies are typically located at high redshift with 
very faint optical counterparts. As a result, only a small fraction of radio sources,
typically located in the local universe, 
have measurements of photometric and spectroscopic redshifts  \citep[e.g.,][]{pn91,M04}.
The only viable approach
towards studying the LSS with radio continuum data is therefore via angular clustering. 
Despite its limitations, such two-dimensional (2D) clustering analyses can be very useful to 
identify the nature of radio sources, probe their evolution, and reveal 
subtle observational systematic errors.

The angular correlation properties of wide-angle radio catalogs have 
long been detected and analyzed both in configuration and in harmonic space, usually using two-point statistics.
The two-point angular correlation function (ACF) of the radio sources has been studied in several of the above-mentioned wide-angle radio samples \citep[e.g.,][]{Cress1996,loan97,Blake2002,Overzier2003,Blake2004_SUMSS,Negrello2006,Chen2016}.
As for the harmonics analysis, the main catalog used to measure the
angular power spectrum (APS) so far is the NVSS \citep[e.g.,][]{Blake2004_NVSS,Nusser2015}.

The results of these analyses have shown that the clustering properties of radio sources
can be accounted for in the framework of Lambda cold dark matter ($\Lambda$CDM) and halo models, in which 
radio sources are located in massive dark matter halos, typically associated to large elliptical galaxies and active galactic nucleus (AGN) activity, sharing a common cosmological evolution.
One exception to this success is represented by the dipole moment in the distribution of the radio sources in the NVSS and other radio surveys.
After its first detection \citep{BW02}, it was clear that the dipole direction agrees with that of the cosmic microwave background (CMB) dipole. However, several subsequent analyses indicated that its amplitude is larger than expected.
The tension with the CMB dipole and theoretical predictions has been quantified 
by \cite{SI11,GH12,RS13,FC14,TJ14,T14} to name a few examples. 
All of these studies agree that the observed dipole is difficult to reconcile with  
the predictions of the standard cosmological model \citep{EB84}, although the significance of the mismatch depends on the 
analysis and can be partially reduced by taking into account the intrinsic dipole in the local LSS \citep{FC14,TN16,C17} or by pushing the analysis to the quadrupole and octupole moments \citep{TA18} .

New, large, homogeneous datasets at different radio frequencies are clearly welcome to investigate the clustering properties of the radio objects in more depth. This is one of the reasons why the TIFR GMRT Sky Survey (TGSS)  at 150 MHz, carried out at the GMRT\footnote{\url{http://www.gmrt.ncra.tifr.res.in/}} 
has received so much attention.
\cite{Rana2018} studied the clustering properties of this sample 
in configuration space by measuring its ACF. Their analysis, which is focused on angular scales larger than $\theta = 0.1^{\circ}$, confirms that in this range the 
 ACF is well described by a single power law with a slope comparable with that of NVSS 
 but a larger amplitude. In another work,
\cite{Bengaly2018} investigated the TGSS clustering properties
in the harmonic space, focusing on the much-debated
dipole moment. Quite surprisingly, they showed that the TGSS dipole is 
also well aligned with that of the CMB,
but that its amplitude is large, much larger in fact that the one observed in NVSS.

The main goal of our work is to expand the analysis of \cite{Bengaly2018} by considering
the full TGSS angular spectrum and compare it with theoretical expectations, focusing on the large-scale behavior. As previous APS models have adopted simplifying hypotheses and
neglected theoretical uncertainties, we shall emphasize the modeling aspects by including 
all the effects that contribute to the clustering signal and by propagating the 
uncertainties on the nature, redshift distribution, and bias of the 
radio sources into the APS model. We aim at quantifying possible departures from 
$\Lambda$CDM on all scales, using all multipoles $\ell>1$.

The outline of the paper is as follows. In Section \ref{sec:data}, we briefly describe the datasets used in this work. These include the TGSS survey that constitutes the 
focus of our research, the NVSS survey that we mainly use as a control sample,
a catalog of radio sources obtained by cross-matching TGSS with NVSS objects,
which we use to identify systematics and, in addition, a sample of 
quasars extracted from the Sloan Digital Sky Survey (SDSS) spectroscopic catalog, to trace the distribution of TGSS objects
at large redshifts.
In Section \ref{sec:clustering_analysis} we present the result of our analysis in configuration (i.e., the ACF) and harmonics (i.e., the APS) space. The motivation for 
considering the ACF is to assess its behavior on angular scales 
smaller than those explored by \cite{Rana2018} in order to isolate and characterize the clustering signal generated by multiple-component radio sources. That section also features the various tests performed to assess the robustness of the results.
The model APS is presented  in Sect. \ref{sec:modeling} and the results of its comparison with the measured TGSS power spectrum are presented in Sect. \ref{sec:chisq analysis}. Our conclusions are discussed in Sect. \ref{sec:conclusion}. 
Finally, in the Appendix we expand the tests performed in Sect. \ref{sec:clustering_analysis}
to search for systematic errors in the TGSS dataset that could potentially affect our APS estimate.

Throughout the paper we assume a flat $\Lambda$CDM cosmological model with parameters taken from \cite{Planck2016}: 
Hubble constant $H_0 = 67.8\,$ km s$^{-1}$ Mpc$^{-1}$, a total matter density parameter $\Omega_m = 0.308$, baryonic density parameter $\Omega_b=0.048$,
the rms of mass fluctuations at a scale of 8 $h^{-1} \mathrm{Mpc}$ $\sigma_{8}=0.815$, and a primordial spectral index $\mathrm{n}_s = 0.9677$.

\section{Datasets}
\label{sec:data}

The main dataset used in this work is the TIFR GMRT Sky Survey (TGSS) of radio objects detected at $150$ MHz. A large fraction of them
are in common with those in the NRAO VLA Sky Survey (NVSS). We analyze both the NVSS as well as the catalog of common objects (dubbed TGSS$\times$NVSS). These radio datasets are employed for angular clustering measurements.
We also use the quasar catalog from SDSS Data Release 14, but only to probe the redshift distribution of the TGSS sample.

\subsection{The TGSS catalog}

TGSS\footnote{\url{http://tgss.ncra.tifr.res.in}} is a wide-angle continuum radio survey 
at the frequency of 150 MHz, performed with the 
GMRT radio telescope \citep{Swarup1991}
between April 2010 and March 2012.
The survey covers $36,900\, \mathrm{deg^{2}}$ above $\delta > -53\degr$ (i.e., $\sim \ 90\%$ of sky). 
In this work we use the TGSS Alternative Data Release 1 \citep[ADR1,][]{Intema2017}\footnote{\url{http://tgssadr.strw.leidenuniv.nl/doku.php}}, which is the result of an independent re-processing of archival TGSS data using the SPAM package \citep{Intema2009}.

TGSS ADR1 contains $623,604$ objects for which different quantities are specified. For this work
we use angular positions, as well as the integrated flux density at 150 MHz and its uncertainty. 
The overall astrometric accuracy is better than $2\arcsec$ in right ascension and declination, and the flux density accuracy is estimated to be $\sim 10 \%$.
We shall consider only objects with integrated flux density above $S_{150} = 100\, \mathrm{mJy}$, where the ADR1 catalog is $\sim \, 100\%$ complete
and more than $99.9\%$ reliable (fraction of detections corresponding to real sources, \citealt{Intema2017}).
The resolution of the survey depends on the declination:  
it is $25\arcsec \times 25\arcsec$ north of $\delta \sim 19\degr$ and $25\arcsec \times 25\arcsec / \cos(\delta - 19\degr)$ south of $\delta \sim 19\degr$.

\begin{figure}
\vspace{-1.5cm}
\includegraphics[width=\columnwidth]{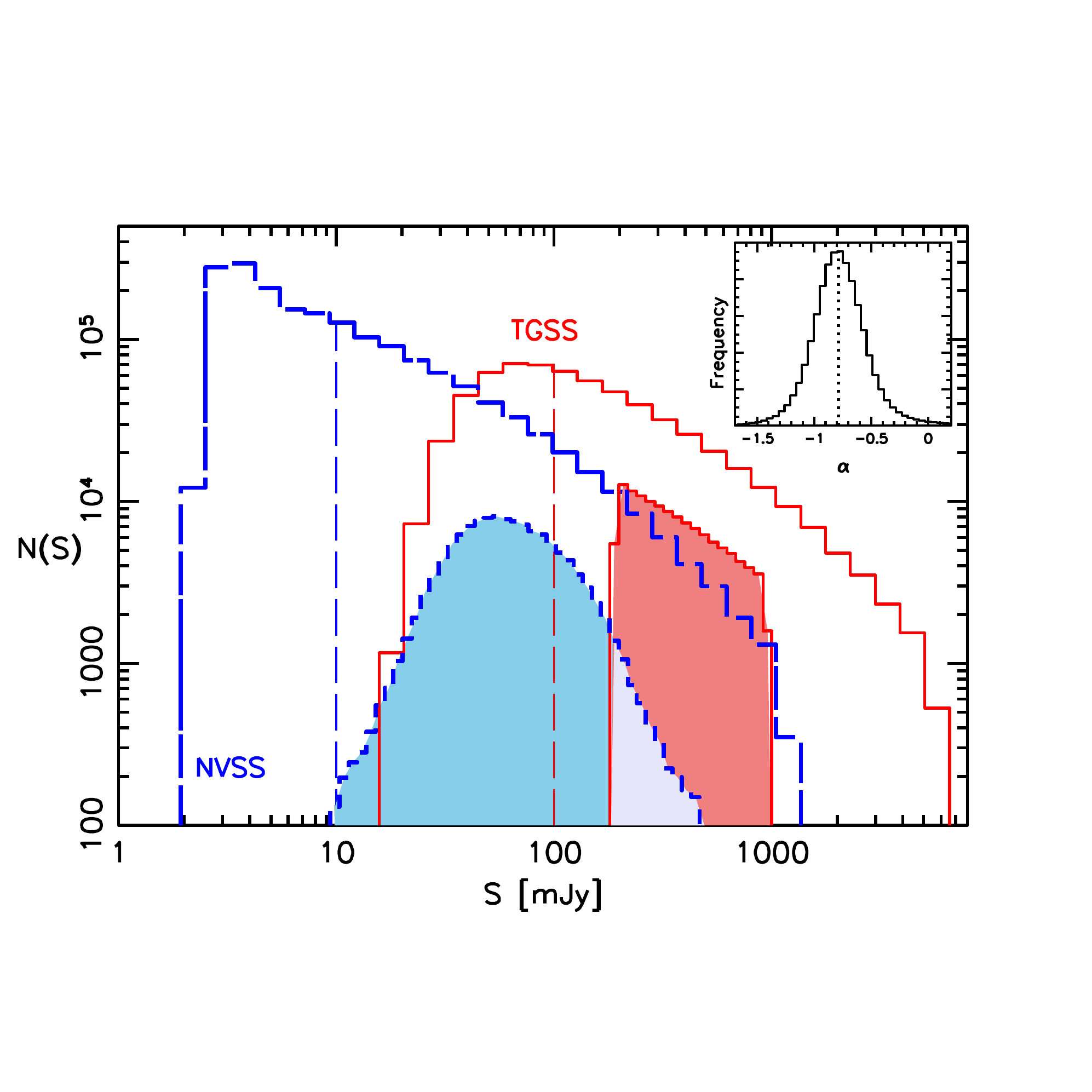}
\vspace{-1.8cm}
\caption{Source counts of the TGSS (red, continuous) and NVSS (blue, long-dashed) catalogs. 
The red shaded histogram on the right shows the number counts (in $S_{150}$ flux unit) 
of the objects in the TGSS$\times$NVSS catalog. The blue shaded area on the left 
shows the number counts (in $S_{1.4}$ flux unit) of the same TGSS$\times$NVSS objects. 
Dashed vertical lines indicate the lower flux thresholds assumed for the analysis presented in this paper. The histogram in the 
insert shows the distribution of the 150 MHz - 1.4 GHz spectral index  of the sources in the TGSS$\times$NVSS catalog. The vertical dotted line indicates the peak of the distribution at {\bf $\alpha=-0.77$}.}
    \label{fig:TGSS_ADR1_flux_distr}
\end{figure}

The red histogram in Fig.~\ref{fig:TGSS_ADR1_flux_distr} shows the TGSS source counts $N(S)$ per logarithmic flux bin ($\Delta \log(S) = 0.114$) per solid angle. The turnover at $S_{150} \sim 70\, \mathrm{mJy}$ reveals the completeness limit of the survey and justifies our 
conservative choice of considering only objects that are brighter than $100\, \mathrm{mJy}$. Beyond this flux the $N(S)$ is well fitted by a power law that, as pointed out by \cite{Bengaly2018}, has a slope $S^{-0.955}$  in the range $100\, \mathrm{mJy} < S_{150} < 500\, \mathrm{mJy}$. At brighter fluxes the $N(S)$ becomes steeper.

For our analysis we extract a subsample of the TGSS objects. The main selection criterion 
is the noise level, which is not constant across the survey (the median {\it RMS} value is $3.5\, \mathrm{mJy/beam}$); it increases towards the Galactic plane and near bright radio sources.
The TGSS subsample used in this work was selected as follows:

\begin{itemize}
\item[•] We exclude all objects with declination  $\delta < -45\degr$, where the {\it RMS} noise is higher than $\sim 5\, \mathrm{mJy/beam}$, which is the value below which $80\%$ of all measurements lie \citep[see,][Fig. 7]{Intema2017}.

\item[•] We discard objects with Galactic latitude $|b_\mathrm{Gal}| < 10\degr$, where the {\it RMS} noise is also large due to bright diffuse synchrotron emission of the Galaxy and to the presence of Galactic radio sources. 

\item[•] We discard the sky patch of coordinates $97.5\degr < \alpha < 142.5\degr$ and $25\degr < \delta < 39\degr$, corresponding to the problematic observing session on January 28, 2011
characterized by bad ionospheric conditions \citep{Intema2017}. 

\item[•] Following visual inspection using the Aladin Desktop tool, we mask out 34 of the brightest extended radio sources 
that appear as a cluster of many points in the catalog which could produce anomalous large counts in small regions, mimicking spurious small-scale clustering \citep{Nusser2015}.

\end{itemize}
The areas of the sky identified by these constraints are represented by a binary \texttt{Healpix} \citep{Healpix} mask with resolution $N_{\rm side}=512$, which corresponds to a pixel size of $0.114\degr$ ($\sim 7\arcmin$) or a pixel area of $0.013 \, \mathrm{deg^{2}}$. 
The maximum multipole corresponding to this angular resolution is {\bf $\ell_{max} \simeq 1024$}. Nevertheless, in our analysis we only consider modes $\ell<100$, to minimize nonlinear effects, as detailed in Sect. \ref{sec:clustering_analysis}. After applying this mask, the fraction of the sky covered by the TGSS catalog 
is $f_\mathrm{sky} \simeq 0.7$. Both very bright and very faint sources have also been excluded.
Since different subsamples are considered for the clustering analyses, here we only specify the least restrictive flux cuts, that define the largest sample considered, and those used to extract the TGSS sample that we use as a reference; referred to as the \textit{Reference} sample. The other flux cuts will be specified 
in Section \ref{sec:fcuts}, where they are used.

\begin{itemize}

\item[•] We exclude all objects brighter than $5000\, \mathrm{mJy}$ since they increase the {\it RMS} noise in localized regions and produce spurious clustering signal.
This threshold corresponds to the flux cut of about $1000\,
\mathrm{mJy}$ in the 1.4-GHz band that we have adopted for the NVSS sample 
(see following section). For the {\it Reference} TGSS sample we set a more conservative 
conservative flux cut $S_{150}= 1000\, \mathrm{mJy}$ to minimize the chance of systematic effects
that, as shown below, have a more significant impact than the random sampling noise.
However, we demonstrate in Sect. \ref{sec:fcuts} that the results of our analysis are very robust to the choice of the upper flux limit, in particular when this is set equal to $5000\, \mathrm{mJy}$.

\item[•] Similarly, as already mentioned, we exclude all objects fainter than the completeness limit of $S_{150}=100\, \mathrm{mJy}$, but in our {\it Reference} sample we use a stricter lower cut $S_{150}=200\, \mathrm{mJy}$.

\end{itemize}

To summarize, we have defined a TGSS {\it Reference} catalog of $109,941$ radio sources with fluxes in the range $S_{150}=[200,1000] \, \mathrm{mJy}$ located outside the masked area defined above. The main properties of this sample, together with two others used in the analysis (see below), are provided in Table \ref{tab:dataset_params}.

\begin{table}
\center
\begin{tabular}{|c|c|c|c|c|} \hline \hline
Sample & N. objects & $f_{\rm sky}$ & Shot noise  & $\Delta C_{\ell} \times 10^{6}$ \\ \hline
Ref. TGSS & 109,941 & 0.7 & $8.01\times10^{-5}$ & $8.51 $\\
Ref. NVSS & 518,894 & 0.75 & $1.82\times10^{-5}$ & $2.55 $ \\
TGSS$\times$NVSS & 103,047 & 0.67  & $8.23 \times 10^{-5}$ & $8.51 $ \\
    \hline \hline
\end{tabular}\caption{Main datasets used in this work and their characteristics. 
Col. 1: Dataset name. Col. 2: Number of objects. Col. 3: Fraction of the unmasked sky. Col. 3: Shot Noise. Col. 4: APS correction for multiple sources in units $10^{-6}$. 
 }
\label{tab:dataset_params}
\end{table}

\begin{figure*}
\includegraphics[width = \columnwidth]{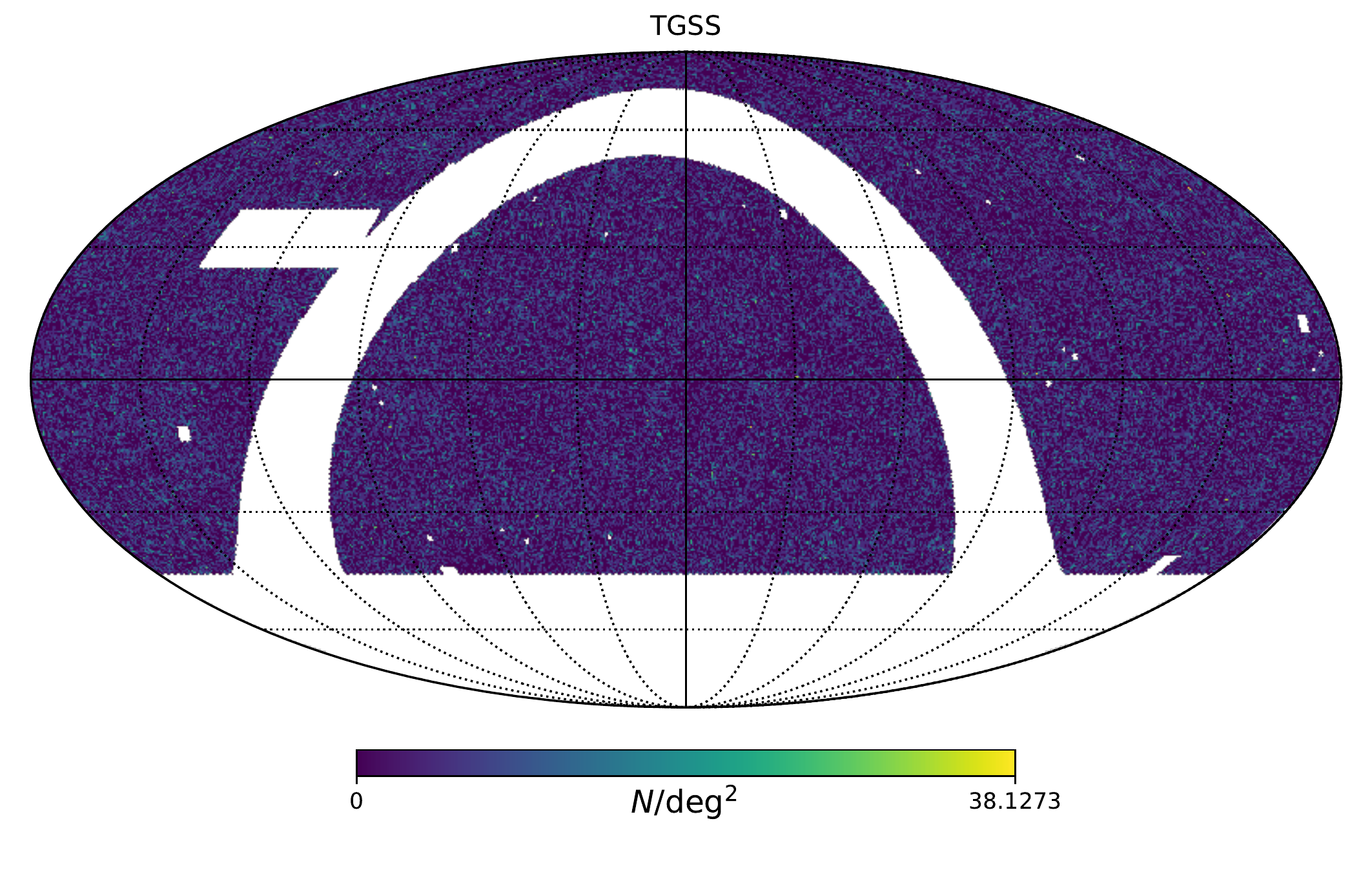}
\includegraphics[width = \columnwidth]{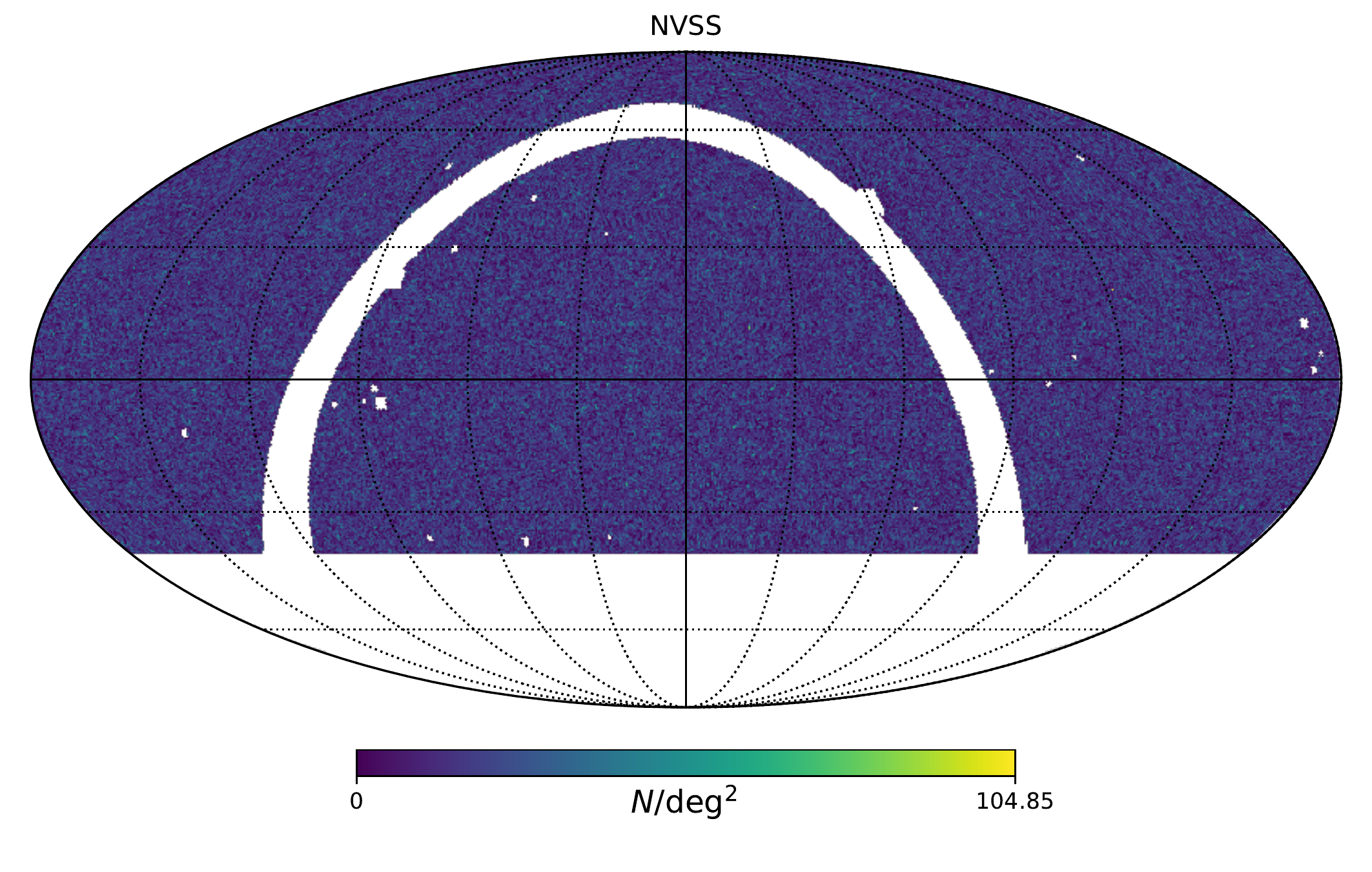}
\caption{Mollweide projection of TGSS (left) and NVSS (right) samples in equatorial coordinates. The plots show reference catalogs with selection criteria described in the text. The color code in the bottom bar refers to $N/\mathrm{deg^{2}}$, denoting the number counts per deg$^{2}$ in the pixel. The resolution of the map is $\mathrm{NSide} = 128$.}
\label{fig:TGSS_NVSS_maps}
\end{figure*}

The left panel of Fig.~\ref{fig:TGSS_NVSS_maps} shows a Mollweide projection of the observed TGSS counts in equatorial coordinates. The color code 
indicates the number counts per squared degree, $N/\mathrm{deg^{2}}$, of TGSS objects with flux in the range $200\, \mathrm{mJy} < S_{150} < 1000\, \mathrm{mJy}$. The masked areas are plotted in a uniform white color.

\subsection{The NVSS catalog}

The 2D clustering properties of the NVSS sources, especially their dipole moment, have been investigated in a number of works. The reason for repeating such an analysis here is twofold.
First of all, it constitutes a useful cross-check for the analogous analysis of the TGSS catalog.
The second, more compelling reason is that, as we see below, the large majority of TGSS sources 
are also listed in the NVSS catalog. Comparing the clustering properties of this population 
with those of their parent catalogs is a useful tool to spot systematic effects and to check 
the robustness of the results to the selection criteria.

The NVSS survey \citep{NVSS} at $1.4\, \mathrm{GHz}$  
contains $\sim \, 1.8$ million sources over an area similar to that of TGSS
and is $99\%$ complete above $S_{1.4} = 3.4\, \mathrm{mJy}$.
Previous works have used  various selection criteria and, consequently, analyzed 
slightly different NVSS samples.
Our data cleaning is similar to that of \cite{Blake2004_NVSS}:

\begin{itemize}
\item[•] We ignore the low signal-to-noise-ratio (S/N) region with declination $\delta < -40\degr$.

\item[•] We exclude objects near the Galactic plane $|b_\mathrm{Gal}| < 5\degr$, to minimize spurious contribution of Galactic foreground and radio sources.

\item[•] We mask out $22$ square regions around bright extended radio sources that can be fitted by multiple elliptical Gaussians and would generate spurious clustering signal \citep{Blake2004_NVSS}.

\end{itemize}

We create a binary \texttt{Healpix} map to quantify the masked region.
After masking, the sky fraction covered by NVSS is $f_\mathrm{sky} \simeq 0.75$.  
Similarly to the TGSS case, we define a {\it Reference} NVSS catalog using the additional flux cuts:

\begin{itemize}

\item[•] A lower cut at $ S_{1.4} = 10\, \mathrm{mJy}$, since below this limit the 
surface density of NVSS sources suffers from systematic fluctuations \citep{Blake2004_NVSS}.

\item[•]  An upper cut at $ S_{1.4} = 1000\, \mathrm{mJy}$, since brighter sources may be associated to extended emission.

\end{itemize}

Our reference NVSS catalog then consists of $518,894$ radio sources with fluxes in the range
$S_{1.4}=[10,1000] \, \mathrm{mJy}$ outside the masked area.
Its source counts are represented by the blue histogram in Fig.~\ref{fig:TGSS_ADR1_flux_distr}. 
Above the 10-mJy threshold (vertical dashed line) the shape of the distribution is similar to that of the TGSS and can be superimposed to it by assuming a TGSS versus NVSS flux ratio
$S_{150}/S_{1.4} \simeq 5$ \citep{Bengaly2018}. 

The right panel of Fig.~\ref{fig:TGSS_NVSS_maps} shows the surface density of NVSS sources outside the masked areas.
It is worth noting that the footprints of the two surveys are very much alike. 
This means that the effects of the two masks 
are very similar and the APS
measured in the two samples can be compared directly.

\subsection{The cross-matched TGSS $\times$ NVSS catalog}

A detailed analysis of the properties of the objects in common between the TGSS and the NVSS
was performed by \cite{Tiwari2016} and \cite{deGasperin2018}. Our goal here is simply to build a matched catalog to estimate the spectral index $\alpha_{\nu}= -1.03 \log(S_{150}/S_{1.4}) $ of the sources in common and
 to investigate their clustering properties in comparison to those of the parent catalogs.

Our procedure of matching the two datasets is as follows:
\begin{itemize} 
\item[•] we consider a TGSS source;
\item[•] we search for NVSS sources within $45\arcsec$ radius, corresponding to the NVSS survey resolution;
\item[•] if a single NVSS source is found, we accept the NVSS object as the cross-match with TGSS;
\item[•] if more than one NVSS source is found, we take the closest one as the cross-match.
\end{itemize}

The resulting cross-matched TGSS$\times$NVSS catalog contains $103,047$ sources within the reference TGSS and NVSS flux limits, corresponding to $\sim 94 \%$ of the TGSS parent sample.
The typical separation between the NVSS and TGSS sources is $1.2\arcsec$ and less than 10 \% of them are separated by more than $8\arcsec$, comparable to the astrometric accuracy, as expected for genuine matches.

The number counts of the TGSS $\times$ NVSS objects are shown in Fig.~\ref{fig:TGSS_ADR1_flux_distr} in both $S_{150}$ (shaded red histogram on the right) and  $S_{1.4}$
(shaded blue histogram on the left) flux units. The counts distribution, characterized by sharp cuts
in  $S_{150}$ flux, has a distribution close to log-normal in units of $S_{1.4}$ flux.

The distribution of the spectral index $\alpha_{\nu}$ is shown in the upper insert of Fig.~\ref{fig:TGSS_ADR1_flux_distr} and is close 
to a Gaussian, with a peak at $ \alpha_{\nu}\simeq -0.77$,
in agreement with previous results \citep{Tiwari2016,deGasperin2018,Rana2018}.

\subsection{The cross-matched TGSS$\times$SDSS-QSO sample}
\label{sec:SDSS}

The last catalog considered here was obtained by cross-matching TGSS sources with the 
quasar (QSO) sample of the Sloan Digital Sky Survey Data Release 14 \citep[SDSS DR14,][]{SDSS_QSO_DR14}.
We point out that,
unlike for the other catalogs described above, we do not expect the TGSS $\times$ SDSS-QSO sample to be statistically representative and, for this reason, we do not use it
to perform any clustering analyses.
Instead, it is only employed to show that the redshift distribution, $N(z)$, of TGSS sources extends out to large redshifts. The motivation is that so far the $N(z)$ of TGSS objects has been estimated directly only at relatively low redshifts by cross-matching them with the galaxies of the SDSS spectroscopic sample \citep{Rana2018}, which do not reach beyond $z=1$.

The observed and model luminosity function of the radio sources \citep{willott}
suggests however that the distribution of TGSS objects should extend to much higher redshifts than SDSS galaxies, so it is worth checking directly that this is indeed the case.
Indirect verification of this prediction already exists: \cite{Nusser2015} cross matched the NVSS catalog with two small spectroscopic surveys (CENSORS and Hercules, \citealt{CENSORS,Hercules}) and found that the distributions of NVSS sources extends out to $z \simeq 3$.  

\begin{figure}
\vspace{-1.8cm}
\includegraphics[width=\columnwidth]{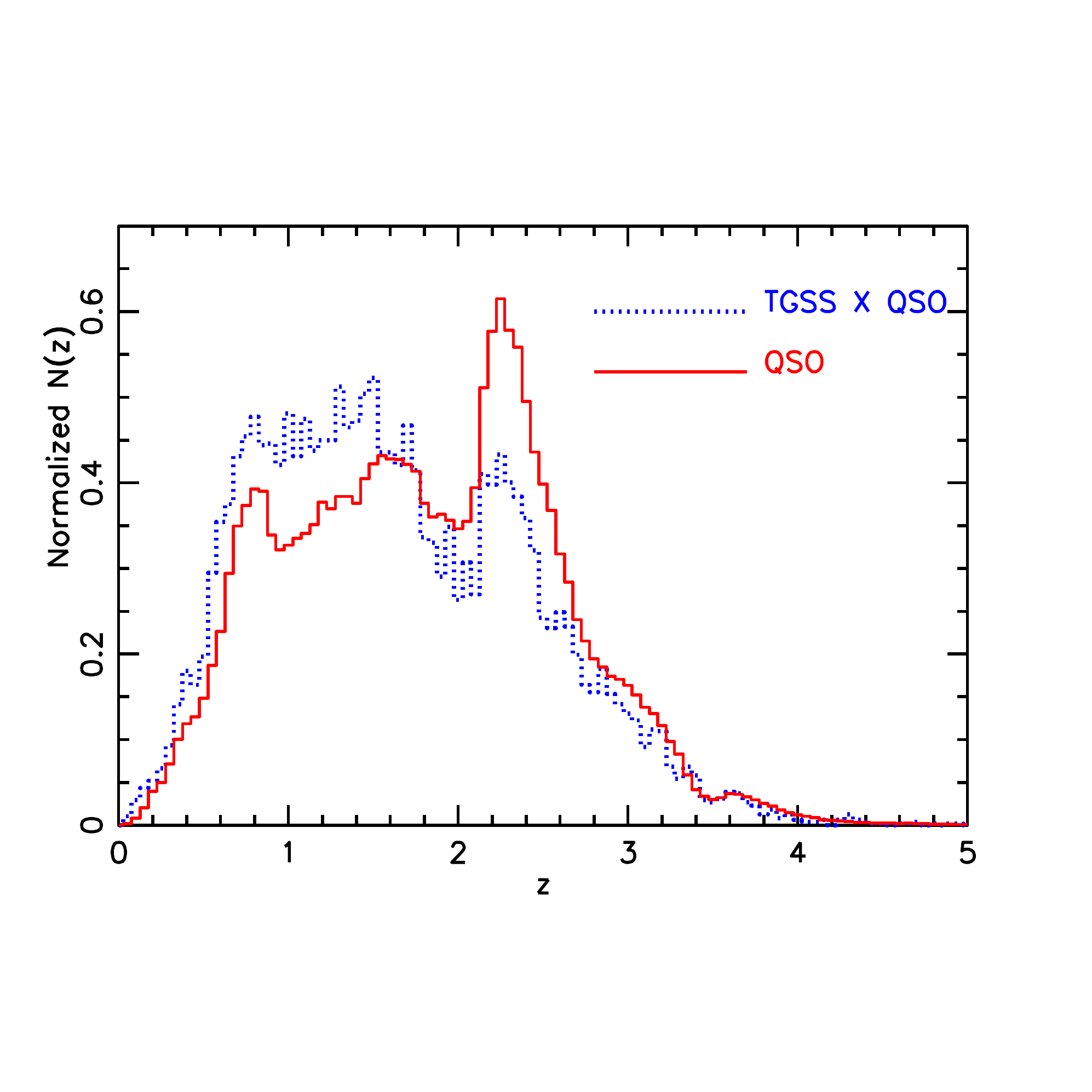}
\vspace{-2cm}
\caption{Normalized redshift distribution of the cross-matched TGSS$\times$QSO catalog (blue, dotted histogram) and of the parent SDSS-DR14 QSO catalog (red, continuous).}
    \label{fig:TGSS-SDSS_DR14Q_cross-match_redshift_distribution_plot_number_of matches_9645}
\end{figure}

To prove that this is also the case for the TGSS sources, we
performed a similar matching procedure as described above to build a TGSS$\times$SDSS-QSO cross-matched catalog.
In the process we ignored astrometric errors in the QSO positions, which are negligible, and searched for TGSS - QSO matches within the angular resolution of TGSS ($25\arcsec$) and in the  area common to the two surveys.
We found $9,645$ matches corresponding to $\sim 1.5\%$ of the TGSS sources, most of them within $8 \arcsec$ of the target object. The fraction of matched objects is small but still considerably larger than that of objects with an optical counterpart
in the SDSS galaxy catalog ($\simeq 0.6\%$, \citealt{Rana2018}). 
The distribution of the TGSS$\times$SDSS-QSO sample extends to $z\sim4$ (blue, dotted histogram in Fig.~\ref{fig:TGSS-SDSS_DR14Q_cross-match_redshift_distribution_plot_number_of matches_9645}), which is far beyond the redshift probed by \cite{Rana2018}, and is characterized by a double peak
like that of the parent DR14 QSO sample (red, continuous histogram in Fig.~\ref{fig:TGSS-SDSS_DR14Q_cross-match_redshift_distribution_plot_number_of matches_9645}),
suggesting that this small cross-matched catalog traces the redshift distribution of 
the optically selected QSO population.

The fact that this redshift distribution is so different from the one found by \cite{Rana2018}
strongly suggests that the TGSS catalog contains various types of radio sources. We shall take into account this fact to model their correlation properties.

\section{Clustering analysis}
\label{sec:clustering_analysis}
In this section we describe the statistical tools and the main results 
of the TGSS clustering analysis.
We mainly use the angular power spectrum. However, the auto correlation function is also considered.


\subsection{The two-point angular correlation function}
We measure the angular two-point correlation function using the \texttt{TreeCorr} package \citep{TreeCorr2004},  which implements 
the minimum variance estimator of \citet{LS93}. This estimator consists of counting and combining pair counts of genuine objects and random sources distributed within the same surveyed area as the real one, but without intrinsic clustering.
The catalog of random sources contains ten times as many objects as the real catalog, and accounts for the complex geometry of the sample. It does not however correct for possible large-scale gradients induced by systematic uncertainties which therefore need to be identified and accounted for on a case by case basis.
The \texttt{TreeCorr} package generates ACF in bins of width $\Delta \log (\theta(\degr)) = 0.1$,
along with estimated errors obtained from propagating the Poisson noise. 

The cosmic variance contribution could be estimated under the assumption of Gaussian errors from Eq. 20 of \cite{EZ01}. Here we prefer to ignore this term since at the angular separations considered in our analysis ($\theta \leq 0.1 \degr$) the Gaussian approximation is expected to break down and the error budget to be dominated by Poisson noise rather than cosmic variance.
For the same reason we ignore the effect of the ``integral constraint'', that is, the fact that the mean surface density of the sources is computed over a fraction of the sky \citep[e.g.,][]{RE99}.  
Given the large areas covered by the radio samples  and the small angular scales considered here, the integral constraint is small and can be neglected.

\begin{figure}
\vspace{-1.6cm}
\includegraphics[width=\columnwidth]{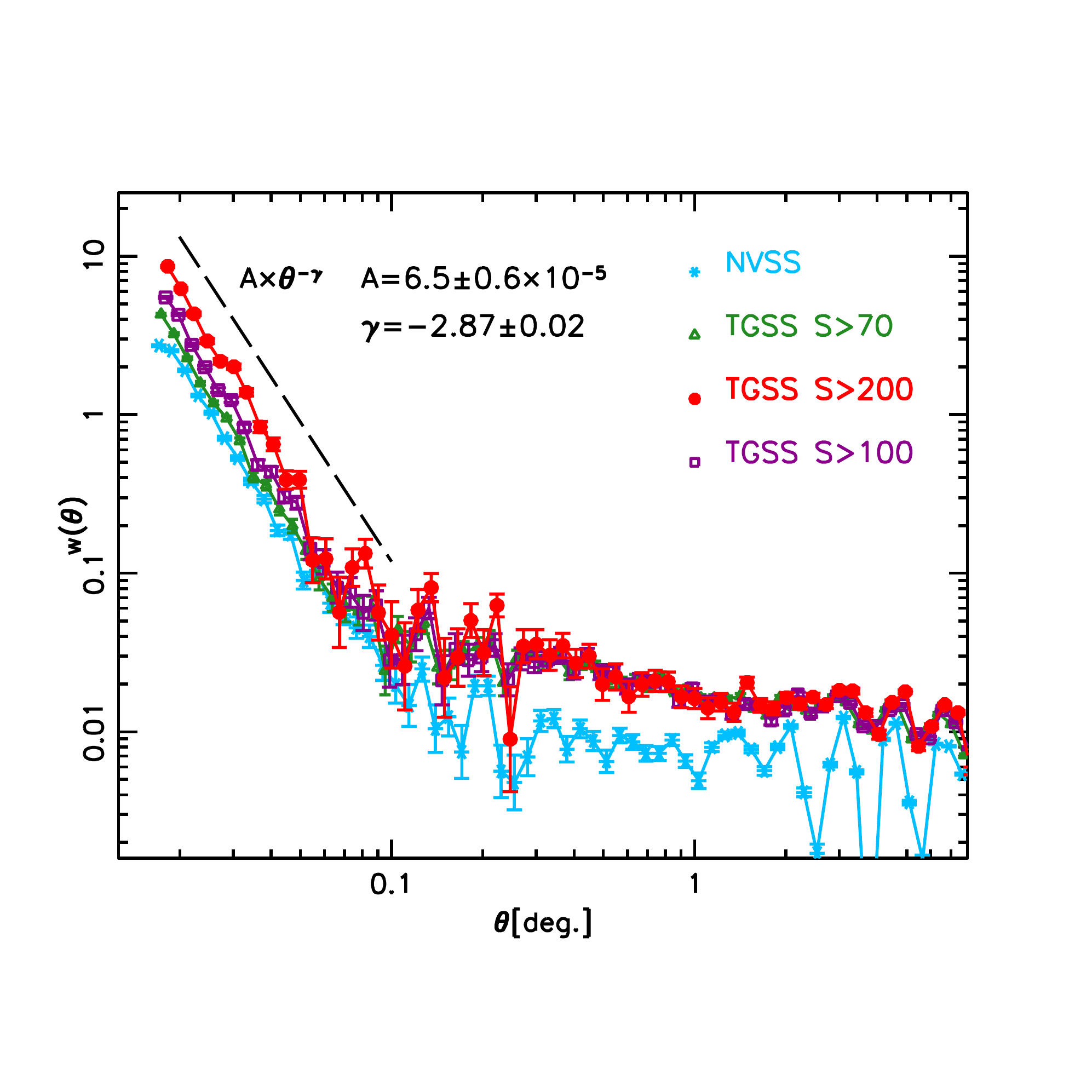}
\vspace{-1.2cm}
\caption{Angular two-point correlation function for the {\it Reference} TGSS (red dots) and NVSS (light blue asterisks) samples.
Green triangles and purple squares represent the ACF of two additional TGSS subsamples selected at different flux cuts 
$S_{150}>70\, \mathrm{mJy}$ and $S_{150}>100 \, \mathrm{mJy}$, respectively. Error bars represent Poisson uncertainties.
    The black dashed line shows the best-fit power law  to the ACF of the reference sample at $\theta < 0.1\degr$. A vertical offset has been applied to avoid overcrowding. The best fitting parameters are indicated in the plot.}
    \label{fig:TGSS_autocorrelation_function_compare_samples}
\end{figure}

Figure \ref{fig:TGSS_autocorrelation_function_compare_samples} shows the measured ACF of the {\it Reference} TGSS catalog (red dots with error bars) and  of the {\it Reference} NVSS catalog (small cyan asterisks).
Both ACFs exhibit a characteristic double power-law shape \citep{Blake2002}
which reflects the fact that while on scales larger than
$\theta \simeq 0.1\degr$ the signal is dominated by the  correlation among sources in different dark matter halos (i.e., the two-halo term), at smaller scales it is dominated by correlation of multiple sources within the same halo (the one-halo term).
This second term depends on the density profile of the source, the typical number of radio components per source and the fraction of sources with multiple radio components.
In the harmonic space, this one-halo term generates an almost constant, shot-noise-like signal  that needs to be accounted for to compare the measured APS with theoretical predictions.
The magnitude of this term depends on the characteristics of the sample: in brighter samples with a larger number of extended sources (and thus with a larger fraction of objects with multiple radio sources) this term is large, which explains why in Fig.~ \ref{fig:TGSS_autocorrelation_function_compare_samples} the amplitude of the TGSS ACF 
increases with the flux threshold. 

Following \cite{Blake2002} we compute this term by fitting a power law to the measured ACF below $\theta=0.1 \degr$, under the hypothesis that the number of radio components per TGSS source  is the same as in  NVSS. In Fig.~ \ref{fig:TGSS_autocorrelation_function_compare_samples} we show the best-fitting power law to the 
reference TGSS sample (dashed line vertically offset to avoid confusion) together with the values of the best-fit amplitude $A$ and slope $\gamma$.
As these are different for the different TGSS subsamples, the best-fitting procedure has been repeated  for all TGSS subsamples considered in our analysis.

For the {\it Reference} TGSS sample we estimate that the fraction of TGSS sources with multiple components is $e=0.09\pm 0.009$,
where errors on $e$ are propagated from the uncertainties of the measured ACF parameters  $A$ and $\gamma$. The corresponding shot-noise-like correction
that we apply to the measured angular spectrum is $\Delta C_{\ell}\simeq 2e\sigma_N / (1+e)=(8.51\pm 0.66)\times 10^{-6}$, where $\sigma_N$ is the surface density.
For the {\it Reference NVSS} case the corresponding values are  $e=0.07\pm 0.005$  and  $\Delta C_{\ell}=(2.55\pm 0.18) \times 10^{-6}$ (see Table \ref{tab:dataset_params}).

\subsection{The angular power spectrum}
\label{sec:APS}
To measure the APS we use the estimator introduced by \citet{peebles}, 
implemented and described in detail by for example \cite{Blake2004_NVSS,2011MNRAS.412.1669T,BABBP2018}. This estimator is based on harmonic decomposition of the observed distribution of galaxies expressed in Healpix maps, and generates  estimates of the APS which are corrected for partial sky coverage and Poisson noise. We focus our analysis 
on the multipole range $2 \leq \ell \leq 100$ and consider the angular power in bins $\Delta \ell=5$.
We neglect the mode $\ell=1$ because the  
dipole of TGSS sources has already been studied by \cite{Bengaly2018}.
The reason for setting $\ell \leq 100$ is to reduce the impact of nonlinear effects that correlate 
modes with large multipoles $\ell$. Mode coupling is also induced by the incomplete sky coverage, although the effect is not expected to be large, given the wide areas of the NVSS and TGSS catalogs.
The $\Delta \ell=5$ bin is introduced to further reduce the effect of mode coupling because the effect of
binning is to decorrelate measurements, resulting in a more Gaussian likelihood \citep{2011MNRAS.412.1669T}.
For all these reasons we assume Gaussian-independent random errors that, for the 
individual $\ell$ mode, can be expressed as \citep[see e.g.,][]{dodelson}:

\begin{equation}
\sigma_{C_{\ell}} = \sqrt{\frac{2}{(2\ell + 1)f_{\rm sky}}} \left(C_{\ell} + S \right) \, ,
 \label{errorbars}
\end{equation}
where $S = \sigma_N^{-1}$ is the Poisson shot-noise contribution
and $f_{\rm sky}$ is the fraction of the unmasked sky covered by the sample (Table \ref{tab:dataset_params}).

\begin{figure}
\vspace{-1.8cm}
\includegraphics[width=\columnwidth]{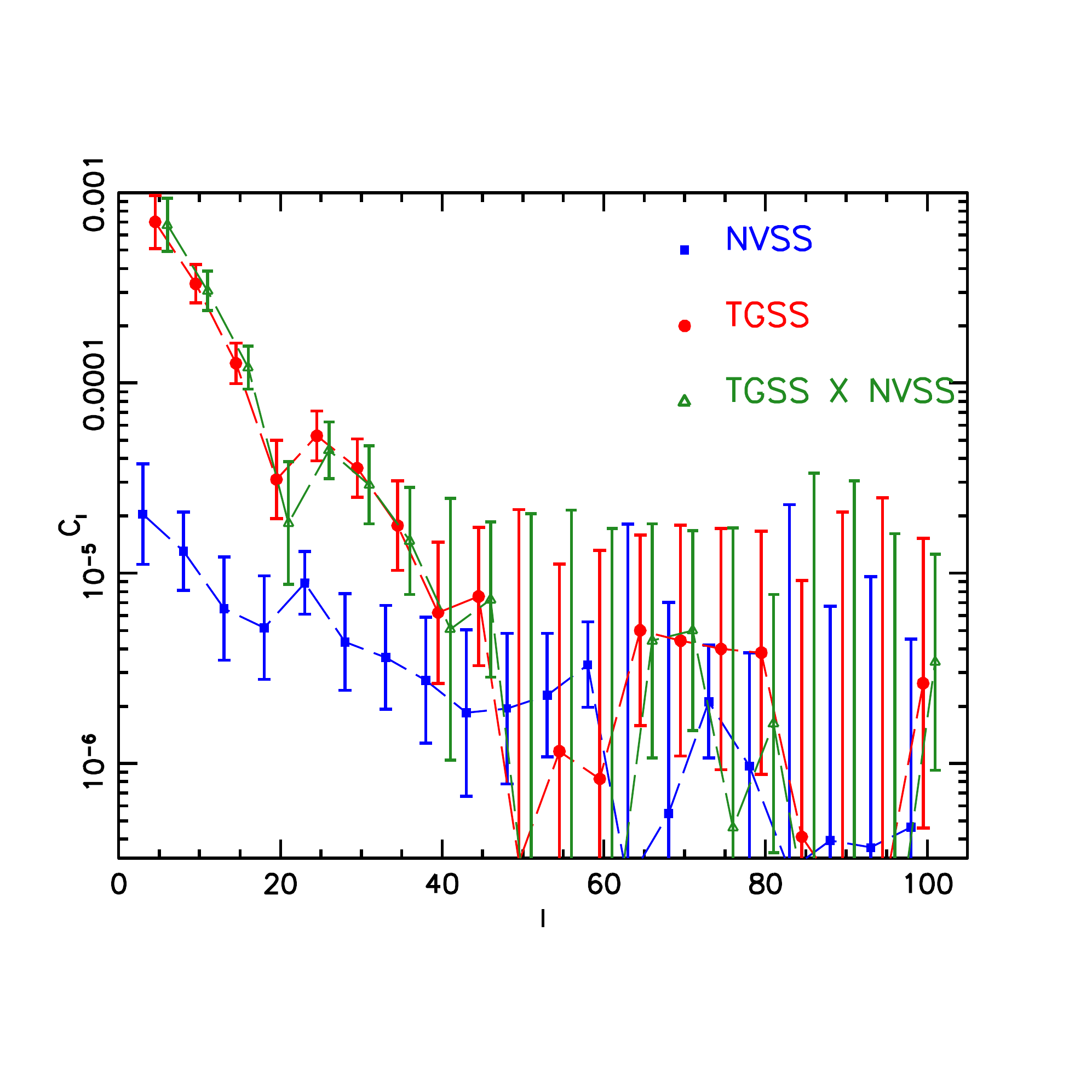}
\vspace{-1.5cm}
\caption{Angular power spectrum of the  NVSS (blue squares) and TGSS (red dots) samples with Gaussian error bars. Small green triangles show the APS of the 
TGSS$\times$NVSS matched catalog.
All spectra are corrected for shot noise and multiple source contributions $\Delta C_{\ell}$.}
    \label{fig:Angular_spectra_TGSSvsNVSS}
\end{figure}

Figure \ref{fig:Angular_spectra_TGSSvsNVSS} compares the measured APS of TGSS (red dots) and NVSS (blue squares) samples, as well as that of the
 TGSS$\times$NVSS cross-matched sample (green triangles).
All spectra are corrected for the multiple source contributions $\Delta C_{\ell}$ 
listed in Table \ref{tab:dataset_params}.
Error bars represent the $1\sigma$ Gaussian uncertainties (Eq. \ref{errorbars}).

The NVSS and TGSS samples considered in the plot are slightly different from the {\it Reference} ones since we applied the same angular mask obtained by multiplying the TGSS and NVSS masks 
pixel by pixel. The sky fraction covered by both samples is 
$f_{\rm sky} \simeq 0.67 $, the same as that covered by TGSS$\times$NVSS. 
The rationale behind this choice is to eliminate all the differences that may result from sampling different regions (cosmic variance) and geometries (convolution
effects).

There is a striking difference between the TGSS and NVSS angular spectra below $\ell \simeq 30$, where the amplitude of the former is significantly larger than that of the latter.
At larger multipoles the two spectra agree with each other within the errors.
The angular spectrum of the matched TGSS$\times$NVSS catalog is similar to that of TGSS-only, which should be expected considering that almost $95\%$ of the TGSS reference sample have counterparts in NVSS.
Taking into account the lack of a one-to-one relation between multipoles $\ell$ and 
angular separations $\theta$, we identify the  amplitude mismatch between the TGSS and NVSS power spectra at $\ell \leq 30$, with the 
 amplitude difference of the angular correlation functions seen at $\theta \geq 0.3\degr$ (Fig.~ \ref{fig:TGSS_autocorrelation_function_compare_samples}).

A useful sanity check to assess the reliability of this result is to 
compare our NVSS APS with the ones measured by \citet{Blake2004_NVSS} and \cite{Nusser2015}.
The test is successful in the sense that it shows a good qualitative agreement with the 
measured spectra in both cases. 
Analogously, it is useful to compare our TGSS angular spectrum with the one computed by 
\cite{Bengaly2018}. Although their analysis focused on the dipole moment,  figure 4 in their paper shows a significant  mismatch between the TGSS and the NVSS APSs at $\ell < 30$
which is analogous to the one detected in our analysis.

This discrepancy between the two spectra is quite unexpected, considering the similarities 
between the two samples both in terms of surveyed areas and the likely nature of the sampled sources.
It may either reflect a genuine physical origin, related to the intrinsic clustering 
properties and redshift distribution of the two samples, or it could be an artifact 
produced by some observational systematic errors that have not been properly identified and accounted for.
In the remainder of this section we explore the latter possibility by  
performing a number of tests aimed at testing the robustness of the APS measurements
to different observational quantities that 
are expected to correlate with the measured radio flux and Galactic emission.

\subsubsection{Robustness to flux cuts}
\label{sec:fcuts}

Spurious clustering features on large angular scales can be generated by 
errors in the flux calibration that are coherent across large areas. 
This type of systematic uncertainty is indeed present and can be significant 
for low-frequency radio observations \citep{aska} reaching up to 
$10-20$ $\%$ in amplitude for the case of the TGSS survey \citep{HW17}. 
The angular scale of coherence is related, in the TGSS case, to the size of the area 
covered during the observing session which is typically of the order of $\sim 10 \degr$
\citep{Bengaly2018}.
The impact of this effect was simulated by \cite{Bengaly2018} who focused on the dipole 
moment, and it turned out to be quite small ($\sim 1$ \% on the dipole amplitude).
This is much smaller than the TGSS versus NVSS power mismatch and can hardly explain it,
even taking into account that its amplitude may increase at $\ell>1$, on the 
angular scales corresponding to those of the typical observational session.
For this reason we exclude this possibility and neglect the effect of 
flux calibration errors in this work.

Other possible systematic errors that are not related to flux calibration can 
be induced by the flux threshold used to select the sample. 
For example, random uncertainties in the flux measurements, that in the TGSS case are of the order of $10 \%$ \citep{Intema2017}, can scatter objects fainter than the completeness limit of the survey into the catalog. Their impact in the APS can be appreciated by changing the 
value of the lower flux threshold $\ubar{S}_{150}$. 
Analogously, including bright, extended  objects associated with multiple sources may artificially increase the clustering signal. In this case, an effective robustness test would be to change the 
upper flux cut $\bar{S}_{150}$ of the TGSS survey.

\begin{figure}
\vspace{-0.8cm}
	\includegraphics[width=\columnwidth]{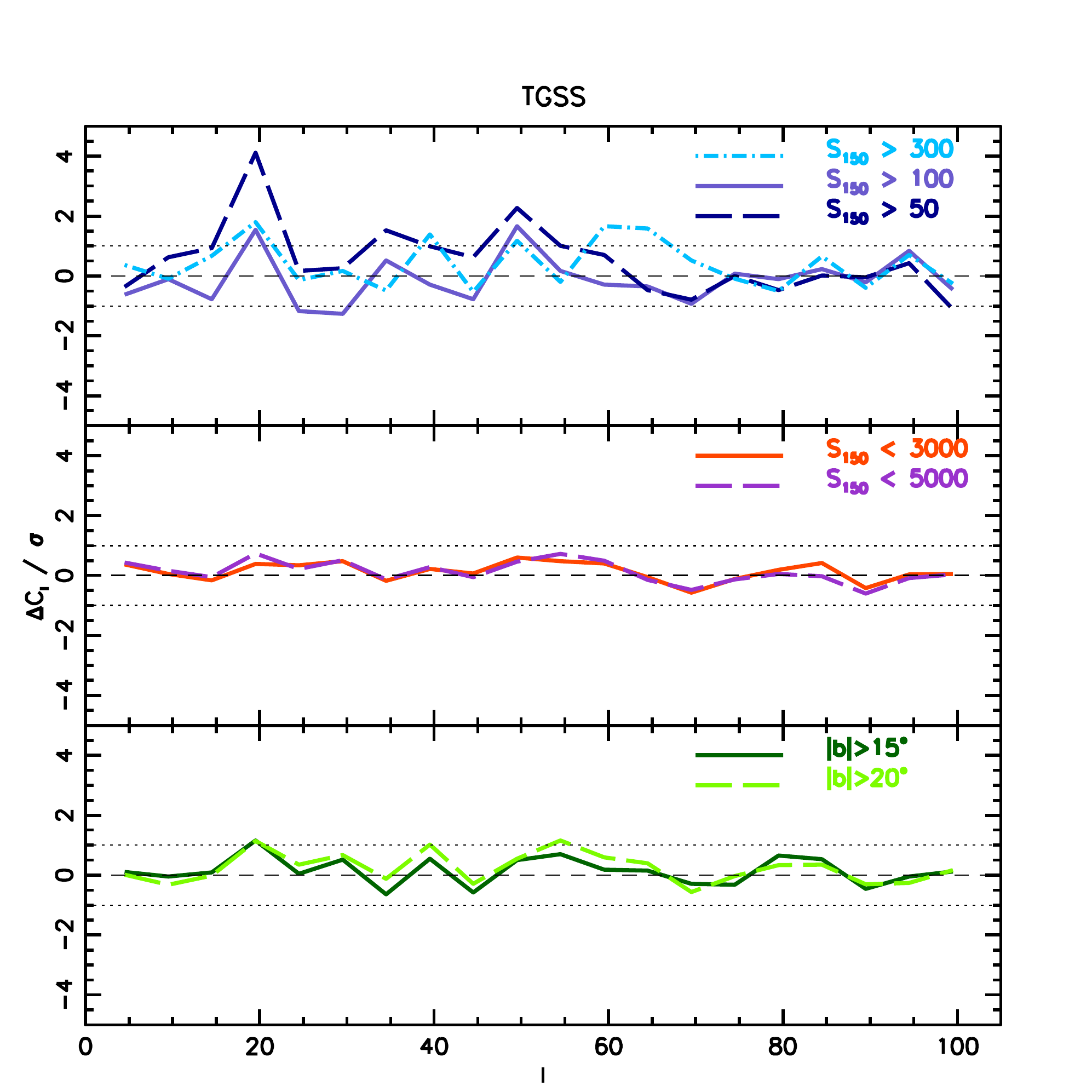}
    \caption{Angular power spectrum residuals of different TGSS samples 
    with respect to the \textit{Reference} case, expressed in units 
    of Gaussian errors. The upper panel shows the normalized residuals of the TGSS samples selected at different values of the minimum flux cut, $\ubar{S}_{150}$, indicated in the plot, compared to the \textit{Reference} case of $ \ubar{S}_{150} = 200 \, \mathrm{mJy} $. 
    In the middle panel we consider samples selected at different values of the maximum flux cut, $\bar{S}_{150}$; the \textit{Reference} is $\bar{S}_{150}  = 1000 \, \mathrm{mJy}$. The bottom panel shows the residuals for samples with different geometry masks, cut at  
different values of the Galactic latitude, also indicated in the plot, referred to the baseline case of $|b|>10^\circ$. 
    The dotted horizontal lines in all panels indicate the 1 $\sigma$ Gaussian error of the \textit{Reference} sample. The dashed horizontal line indicates the zero residual level.}
    \label{fig:APS_robust}
\end{figure}

To quantify the impact of the systematic errors related to the flux cuts 
we ran a set of tests in which the TGSS APS was measured by varying the values of 
$\bar{S}_{150}$ and $\ubar{S}_{150}$, keeping the geometry mask fixed.
The upper panel of Fig.~\ref{fig:APS_robust} illustrates the sensitivity to $\ubar{S}_{150}$.
The curves drawn with different line styles indicate the difference between the APS of the TGSS sample selected at a given cut $\ubar{S}_{150}$ with respect to the {\it Reference} sample, for which $ \ubar{S}_{150} = 200 \, \mathrm{mJy} $.
The difference $\Delta C_{\ell}$ is expressed in units 
of the Gaussian error, $\sigma_{C_{\ell}}$, of the reference APS.
The results are remarkably independent of the choice of the lower flux cut.
Selecting objects with $ \ubar{S}_{150} = 100 \, \mathrm{mJy} $, that is, brighter than the formal completeness limit of the TGSS sample, does not significantly modify the results.
Similarly, when we use more conservative flux cuts of $\ubar{S}_{150} = 300$ and $400 \, \mathrm{mJy}$ (the second one not shown in the plot to avoid overcrowding) we also find results that are consistent with the {\it Reference} ones within the 1-$\sigma$ 
Gaussian errors.

We also tried forcing the lower cut below the TGSS completeness limit, by setting
$\ubar{S}_{150} = 50 \, \mathrm{mJy} $. The rationale behind this choice is to identify possible 
systematic effects that may  be  present also in the complete sample.
We find that using this cut
significantly enhances the power at low multipoles, especially at $\ell \simeq 20$. 
This is a sizable effect that interestingly occurs on the angular scale ($5^{\circ}\times 5^{\circ}$) of the mosaics that constitute the building blocks of the TGSS survey. Since the overall TGSS source catalog is obtained by summing up mosaic-based data, this effect is likely to be attributed to sensitivity variations in adjacent mosaics, or even to the fact that the sensitivity pattern in these mosaics is replicated in adjacent mosaics. As a consequence, the surface density of faint objects with fluxes below the completeness threshold coherently varies across each mosaic, generating a spurious clustering signal on the angular scale of the mosaic itself. 
A small excess of power is also seen at $\ell \simeq 20$ if larger $\ubar{S}_{150}$ 
cuts are applied.
However, its statistical significance is much less than in the $ \ubar{S}_{150} = 50 \, \mathrm{mJy} $ case.
This fact corroborates the hypothesis that this excess power 
reflects an observational systematic effects that are corrected for by selecting objects above
the completeness limit of $S_{150} = 100 \, \mathrm{mJy} $.

Our results are also robust to the choice of the upper threshold $\bar{S}_{150}$, as shown in the middle panel of Fig.~\ref{fig:APS_robust}, which compares two more permissive upper flux cuts at 3000 and 5000 mJy against the {\it Reference} of $\bar{S}_{150}  = 1000 \, \mathrm{mJy}$. 

Finally, we performed analogous robustness tests on the TGSS $\times$ NVSS catalog by 
similarly modifying the upper (lower) flux cuts in both samples below (above) the completeness limits. As for the TGSS sample, we find no significant departures from the {\it Reference} angular power spectrum.

Further tests aimed at detecting possible systematic effects in the TGSS sample that may generate spurious clustering signal are presented in the Appendix.

\subsubsection{Robustness to the choice of the geometry mask}
\label{sec:gcuts}

To quantify possible systematic effects induced by Galactic foregrounds or by any other effect
related to the presence of the Galaxy, we tested the impact of using different geometry masks 
characterized by more conservative cuts in the Galactic latitude.
We explored two cases. In the first one we excluded all objects with $|b| < 15\degr$ and in the second one we discard the region   $|b| < 20\degr$. The unmasked sky fraction is consequently  reduced to $f_{sky} \sim 0.61$ and $f_{sky} \sim 0.56$, respectively.
In both cases we considered the same flux cuts as the reference TGSS sample.

We then computed the residuals of the corresponding angular power spectra with respect to the 
TGSS reference case in units of Gaussian error. The results,
displayed in the lower panel of Fig.~\ref{fig:APS_robust}, appear to be robust to the 
inclusion of objects near the Galactic plane. It is worth noticing that some difference in the various spectra is to be expected because different geometry masks are used here. They are obviously small, since they contribute to the plotted residuals.

Relatively nearby sources can generate high-amplitude clustering signal that is not fully accounted for in the modeling.
As we do not have information on the distance of the sources, an effective strategy to  minimize the impact of the nearest ones is to exclude objects near the Supergalactic plane. \cite{TJ14} adopted this approach in measuring the NVSS dipole and found that this cut has a negligible impact on the dipole signal. They conclude that the dipole is largely generated by distant objects. 
In Section ~\ref{sec:dNdzs}, we show that the model redshift distribution of 
TGSS sources (in Fig.~\ref{fig:redshift_distribution_histogram}) does not feature the prominent local ($z<0.1$) peak that, instead, characterize the NVSS one.
Given the lack of a prominent local population of TGSS objects, we conclude that removing
TGSS objects near the Supergalactic plane will likely only increase the shot noise error and, therefore, we decided not to apply additional cuts to the geometry mask.

\section{Modeling the angular power spectrum of TGSS and NVSS}
\label{sec:modeling}

The analyses performed in the previous sections indicate that the APS of the TGSS sources 
is significantly larger than that of the NVSS at $\ell \simeq 30$ and that the mismatch cannot be attributed to known potential sources of 
observational systematic errors.

In this section we consider the alternative hypothesis that the large-scale TGSS power is genuine and reflects the intrinsic clustering properties of the TGSS radio sources. 
To test this hypothesis we compare the measured APS with the theoretical predictions
obtained assuming a Planck $\Lambda$CDM cosmology \citep{Planck2016} 
and physically motivated models for the 
redshift distribution, $N(z)$, and  bias, $b(z)$, of TGSS sources. 
Since we are interested in large scales, we limit our comparison to the range $\ell \leq 100$.
In this comparison we do not try to infer cosmological parameters, as we assume that the 
background cosmological model is well known. 
Instead, we consider various realistic $N(z)$ and $b(z)$ models to investigate whether
the large-scale power of TGSS can be accounted for
within  the known observational and theoretical errors. 
To assess the validity of this approach we perform the same comparison for the NVSS sample.
Only the {\it Reference} TGSS and NVSS samples are employed here.

To model the APS of TGSS sources we use the code \texttt{CLASSgal} \citep{class,didio} which accounts for nonlinear evolution of matter density fluctuations and offers the possibility to include physical effects such as redshift space distortions, gravitational lensing, and general relativistic effects.
Required inputs are the parameters of the underlying cosmological model (given by our fiducial set of parameters), the redshift distributions of the sources and their linear bias.

All our APS models share the same treatment of the mass power spectrum and differ in the choice of  $N(z)$ and $b(z)$. The characteristics of the model
mass power spectrum are described below. We also quantify the impact of the various physical effects that contribute to the clustering signal
by considering the $N(z) \, + \, b(z)$ model  $S^3$-HB described in the following section.

\begin{itemize}

\item[•] 
\noindent \textit{Nonlinear effects.} The nonlinear evolution of mass density fluctuations 
is modeled within the so-called \texttt{HALOFIT} framework \citep{halofit1,halofit2}.
On the scales of interest ($\ell \leq 100$) nonlinear effects are expected to be small, and for this reason have been ignored altogether in some of the previous APS analyses  \citep[e.g.,][]{Nusser2015}. 
To quantify the impact of nonlinear effects we compared the APS predicted with  \texttt{HALOFIT} with the one obtained using linear perturbation theory, using the
redshift distribution and bias of the model $S^3$-HB.
We found that at $\ell = 100$ the nonlinear evolution enhances the angular power by just $\sim 0.5\%$.

\item[•] 
\noindent \textit{Redshift space distortions} (RSDs). Peculiar velocities amplify the clustering signal on large angular scales. We have  compared the APSs
obtained with and without including RSD and found that RSDs amplify the clustering signal by $\sim 3.5$ \% at $\ell = 2$. The amplitude of the effect decreases at larger multipoles;  it is $\sim 2$ \% at $\ell = 20$ and $\sim 1$ \% at $\ell = 40$.

\item[•] 
\noindent \textit{Magnification lensing.} Gravitational lensing modulates the observed flux of objects and therefore reduces or increases the number counts  above a given flux threshold. This effect generates an additional correlation (or anti-correlation) signal that can be described in terms of magnification-magnification and magnification-density correlations \citep{lensing}.
The magnitude of the effect depends on the slope of the cumulative luminosity function
at the limiting flux of the sample \citep{lensing,didio}. Because of the composite nature of TGSS and NVSS, which contain different types of objects with different luminosity functions (see e.g. below), one needs to account for their individual contributions to the magnification signal.
We do that by considering an effective luminosity function slope that we computed by considering the 
luminosity function of each object type at different redshifts (from \citealt{willott}), 
estimating their slope in correspondence of their limiting flux and computing the effective 
slope as $\tilde{\alpha}={\sum_i\sum_j\alpha(i,j)N_i(z_j)} / {\sum_i\sum_jN_i(z_j)} \simeq 0.3$, where $i$ runs over all object types, $j$ runs over the redshift values, $N_i(z)$ is the redshift distribution of object type $i$ and $\alpha(i,j)$ the slope at the redshift $j$. For this we have assumed the $S^3$-HB model.
We find that in the TGSS case the magnification lensing provides a 
small but significant, negative contribution to the clustering signal.
On the scales of interest ($\ell < 40$) the amplitude of the effect is  $\sim -6\%$,
increasing to  $\sim -9 \%$ at $\ell =2$ and decreasing to $\sim -3 \%$ at $\ell = 100$.

\item[•]
\noindent \textit{General Relativistic effects}. \texttt{CLASSGal} provides the opportunity to include general relativistic contributions to the APS. Their impact, however, is small and limited to very
large angular scales. It is of the order of $1 \%$ at $\ell \sim 4$, sharply decreasing to $0.1 \%$
at $\ell = 30$. 

\end{itemize}

In addition to these physical effects that are included in all our models, there are 
some approximations and corrections that we need to make explicit before 
 considering different model predictions and their comparison with data.

\begin{itemize}

\item[•] 
\noindent \textit{Limber approximation}. Several APS models in the literature have adopted the \cite{Limber} approximation to speed up the APS numerical integration. 
In this work we do not adopt Limber approximation. However, it is useful to quantify 
its impact when comparing our results with those of other analyses.
\texttt{CLASSGal} allows one to switch the Limber approximation option on and off and to select the $\ell$ value above which the approximation is adopted. 
The Limber approximation boosts the modeled angular power at small
 $\ell$ values. In the $S^3$-HB model the
effect is as large as $\sim 15$ \% at $\ell<5$ but then its amplitude rapidly 
decreases to $\sim 7$ \% at $\ell=10$ and to $\sim -1$ \% at $\ell=20$.

\item[•] 
\noindent \textit{Geometry mask}. The effect of the geometry mask is to modulate the signal and to mix power at different APS multipoles.
This effect can be expressed as a convolution of the form 
$\tilde{C}_{\ell }=\sum_{\ell'}R_{\ell \ell'}C_{\ell'}$, where $C_{\ell}$ is the model APS predicted by 
\texttt{CLASSGal} and $R_{\ell \ell'}$ is the mixing matrix, evaluated APS of the survey mask (see e.g. equation 6 of \citealt{BABBP2018}). The main effect of this mask is to modulate power at small multipoles. As we are interested in the range $2 \leq \ell \leq 100$, we do not 
account for the survey beam, which instead would modulate power at large multipoles. 
\end{itemize}

To finalize our APS models of the TGSS and NVSS catalogs we need to specify the redshift distribution and the bias of the sources. 
A specific $N(z) \, + \, b(z)$ model was adopted to quantify the impact of the various effects that contribute to the APS.
Now we describe and justify the adoption of that model and 
explore its uncertainties by considering a number of physically motivated models of both 
$N(z)$ and $b(z)$
that have been proposed in the literature.  We quantify the related theoretical uncertainties by 
taking into account the scatter in the corresponding APS predictions.


\subsection{Redshift distribution models}
\label{sec:dNdzs}

The analysis of the cross-matched TGSS$\times$SDSS QSO catalog has confirmed that
the distribution of TGSS sources extends to much larger redshifts than those
probed by cross correlating them with galaxy redshift catalogs \citep{Rana2018}.
As a consequence, although the 
majority of the TGSS APS signal at low multipoles is probably built up at $z\leq 0.1$ as in the NVSS case \citep{Blake2004_NVSS,Nusser2015}, a non-negligible contribution could also be provided by highly biased objects at higher redshifts.
To test this hypothesis we need to identify the nature of the TGSS 
sources and to probe their distribution along the line of sight.

As we discussed in the introduction, the difficulty in finding IR/optical counterparts 
to the objects identified in low-frequency radio surveys
makes it difficult to measure their $N(z)$ directly.
Only \cite{Nusser2015} have adopted such an approach by cross correlating the NVSS catalog with
a deep but small sample of objects with measured spectroscopic redshifts. With about $300$ matches they were able to trace the redshift distribution of NVSS objects out to $z\sim 3$.
Unfortunately we cannot repeat this procedure with TGSS because of the small number of TGSS matched objects. Therefore we need to change the approach and instead model the TGSS redshift distribution.

For the redshift distribution modeling we use the SKA Simulated Skies ($S^3$) database\footnote{\url{http://s-cubed.physics.ox.ac.uk/}}. This tool, described in detail in \cite{S3}, is meant to 
model radio observations in a given band within a sky patch. It is a phenomenological model 
in the sense that it uses constraints on the available, observed luminosity functions
at different redshifts. 
This simulator also mimics the clustering properties of radio sources by assuming a model for their bias. This latter aspect, however, is quite uncertain, as shown by recent clustering analyses of radio sources \citep{MP17,HJ18}.
In principle, we could have used the newer simulator, T-RECS (\citealt{Bonaldi2019}) that, in addition to predicting more realistic clustering properties than $S^3$, also implements more recent evolutionary models for SFGs, and treats RQ AGNs as part of the SFG class, under the assumption that their radio emission is dominated by star formation.  
However, considering that: {\it i)} we use the simulator to model the redshift distribution of the radio sources and not their clustering properties and {\it ii)} the number of SFGs and RQ AGNs expected in our TGSS sample is negligible, using T-RECS instead of $S^3$ would have little or no impact on our results.
Therefore we decided to stick to $S^3$ instead of using T-RECS that only became available when our work was in a very advanced stage of completion. 

In our application we simulated two radio surveys over the same sky patch of $400 \,\mathrm{deg}^2$ at $150$ MHz and at $1.4$ GHz, and considered objects with fluxes 
above the flux limits of our {\it Reference} samples, that is, $S_{1.4}>10$ mJy and 
$S_{150}>200$ mJy, respectively. No upper flux cuts were considered since, as we have seen, results are very robust to the upper flux cut.   
As a result, we obtained two samples of $\sim 2000$ TGSS-like and $\sim 5000$ NVSS-like sources, respectively.

\begin{figure}
\vspace{-0.5cm}
\includegraphics[width=\columnwidth]{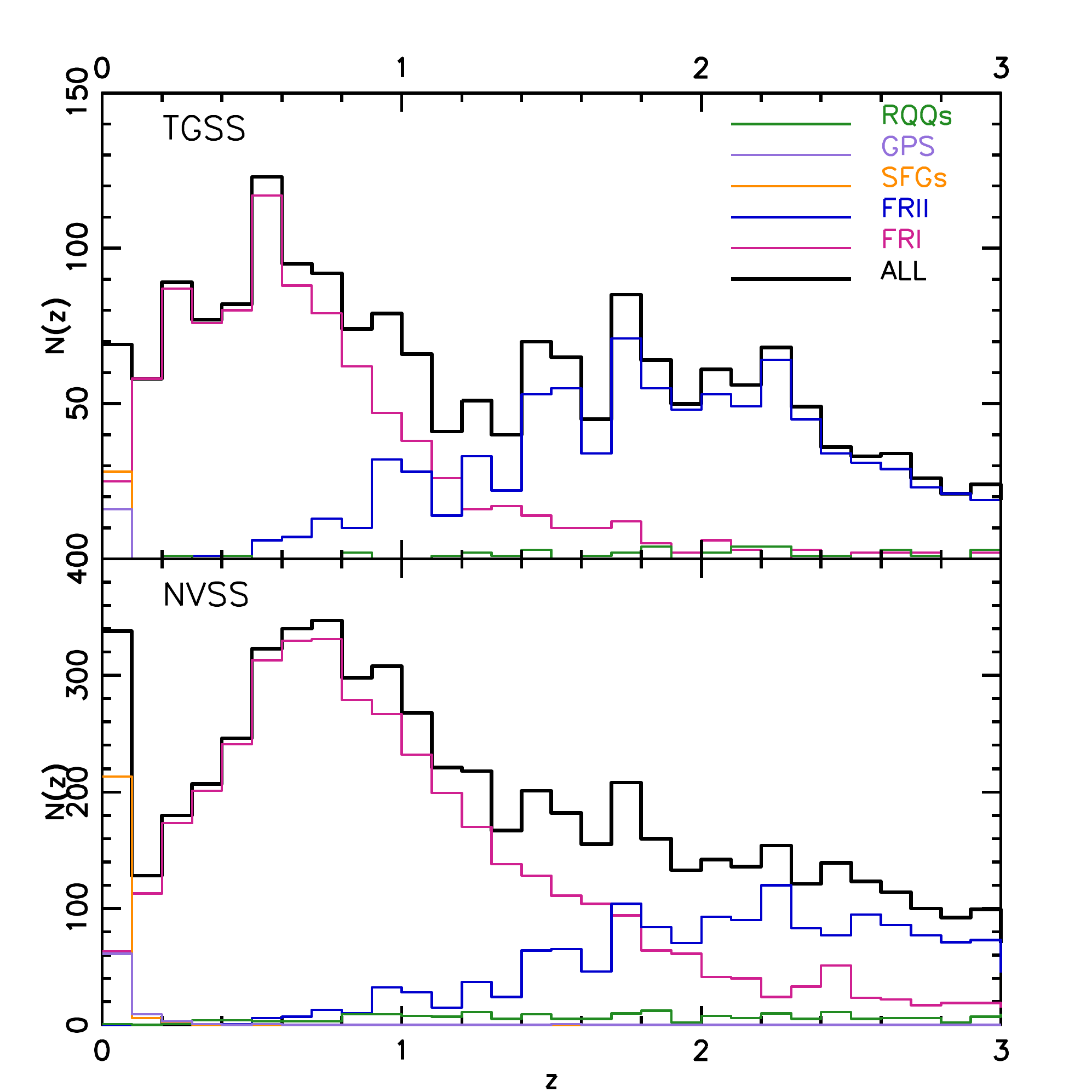}
    \caption{$S^3$ model redshift distributions $N(z)$ of the various types of sources in the TGSS (top) and  NVSS (bottom) samples. The redshift distribution of each source type is represented by a different color, as specified in the upper panel. The thick, black histogram shows the redshift distribution of all types of sources combined.}
    \label{fig:redshift_distribution_histogram}
\end{figure}

The simulator generates five types of radio sources:  {\it i)} star forming galaxies (SFGs), {\it ii)} radio quiet quasars (RQQs), {\it  iii)} Fanaroff-Riley class I sources (FRI), {\it iv)} Fanaroff-Riley class II sources (FRII) and {\it v)} GHz-peaked radio sources (GPSs).
Their redshift distributions in the simulated TGSS and NVSS catalogs are shown 
respectively in the upper and lower panels of
Fig.~\ref{fig:redshift_distribution_histogram} together with the cumulative $N(z)$ (thick line).
In both catalogs the counts are dominated by FRI and FRII-type radio sources. The distribution of FRI objects peaks at $z\sim 0.6$ and dominates the counts at $z<1$. The distribution of FRII objects is much broader and dominates the counts at higher redshifts.
The number of SFGs and GPS objects is much smaller. However, being concentrated in the local universe, they represent a significant fraction of the counts at $z \leq 0.1$. Radio quiet quasars are also comparatively rare and have a very broad distribution, being a sub-dominant population at all redshifts. 

This $N(z)$ model, which we refer to as $S^3$, is the one adopted to predict the APS of
both the NVSS and TGSS samples. It is implemented in the form of a step function 
with the same bin size $\Delta z = 0.1$ used in Fig.~ \ref{fig:redshift_distribution_histogram}.

\subsection{Linear bias models}

The bias of the radio sources is the most uncertain ingredient of the APS model. 
Direct estimates based on cross-matches with CMB lensing convergence maps \citep{allison}, spectroscopic/photometric redshift catalogs \citep{lindsay} 
or by joining the lensing and the clustering information \citep{mandelbaum}
are few, limited to small samples and, therefore, coarsely trace the bias evolution.
In this work in order to appreciate the impact of bias model uncertainties on the 
APS prediction we decided to explore four different, physically motivated, bias models taken from the literature. 
They all assume a deterministic, linear bias that evolves with time (redshift).

\begin{itemize}

\item[•]
\noindent  \textit{Halo Bias model [HB]}. This bias prescription relies on the halo model and assumes that radio sources are hosted in dark matter halos of different masses (and biases). Because of the rarity of radio sources, we assume that halos can host at most one radio source, located at their center. For consistency, we also assume that the radio sample contains the same classes of sources as in the $N(z)$ model. 
We make some hypotheses on the halo host: we adopt the halo bias model, $b_h(M,z)$,
of  \cite{SMTbiasmodel}
and assume that the masses of the halos that host a given source type are Gaussian distributed around a typical mass $\hat{M}$ with a standard deviation $0.2\,\hat{M}$  \citep{ferramacho}. Indicating the 
Gaussian distribution as $G(M,\hat{M})$, we estimate the bias of each type $i$ of radio source as 
\begin{equation}
b_{i}(z)=\int_{0}^{\infty} G_{i}(M,\hat{M}_{i})\,b_{h}(M,z)\, {\rm d}M \, .
 \label{eq:errorbars}
\end{equation}
The values of $\hat{M}_i$  are also taken from \cite{ferramacho}: 
$\hat{M}_{\rm SFG}=1\times 10^{11} M_\odot$, $\hat{M}_{\rm RQQs}=3\times 10^{12} M_\odot$, 
$\hat{M}_{\rm GPS}=\hat{M}_{\rm FRI}=1\times 10^{13} M_\odot$ and $\hat{M}_{\rm FRII}=1\times 10^{14} M_\odot$.

The current implementation of \texttt{CLASSGal} does not allow one to specify different
analytic bias functions $b_i(z)$ for the different source types. To circumvent this problem we approximate each $b_i(z)$ with a step function with the same binning as $N_i(z)$,
and compute the effective {\it bias function} of the catalog:
\begin{equation}
b_{\rm eff}(z)=\frac{\sum_i N_i(z) b_i(z)}{\sum_i N_i(z)} \, .
 \label{eq:beff}
\end{equation}
We then feed the \texttt{CLASS} code with an effective redshift distribution of the objects $\tilde{N}(z)=b_{\rm eff}(z) \times N(z)$. Next we estimate the effective {\it bias parameter} of the sample 
$b_{\rm eff} = {\sum_i \sum_j b_i(z_j)N_i(z_j)} / {\sum_i N_i(z_j)}$ and feed this single parameter to the code as the linear bias of the whole sample.
In Figure \ref{fig:Bias_redshift_evolution_halo_masses} we show the effective {\it bias function} $b_{\rm eff}(z)$ of the NVSS (top) and TGSS (bottom) catalogs for all the models explored and, in particular, for the HB model (purple, continuous curve).

This somewhat cumbersome procedure is analogous to using a normalized redshift distribution
$\hat{N}(z_j)={\sum_i N_i(z_j)} / {\sum_i \sum_j N_i(z_j)}$ and a normalized biasing function $\hat{b}(z_j)= {\sum_i b_i(z_j)N_i(z_j)} / {\sum_i \sum_j N_i(z_j)}$ 
as input parameters to \texttt{CLASSGal}.

\item[•]
\noindent \textit{Truncated Halo Bias  model [THB]}. Some previous analyses (e.g., \citealt{TN16}) have assumed a truncated bias evolution in which the halo bias does not increase 
indefinitely with the redshift but remains constant beyond $z=1.5$, that is, $b_i(z>1.5)=b_i(z=1.5)$. Although this is clearly a rough approximation and there is no compelling theoretical reason to justify an abrupt cut on the bias at high redshift, we also consider this possibility for the sake of completeness and as a robustness test.
In Fig.~\ref{fig:Bias_redshift_evolution_halo_masses} this model is represented by the blue short-dashed curve.

\item[•]
\noindent  \textit{Parametric Bias model [PB]}. \citet{TN16} proposed a parametric bias model for the NVSS sources
also used by \cite{Bengaly2018} to model the TGSS bias. 
The parameters of the parametric models, specified in these works,
have been determined by best-fitting the number counts and angular spectra of the radio sources.

\begin{figure}
\vspace{-1.6cm}
\includegraphics[width=\columnwidth]{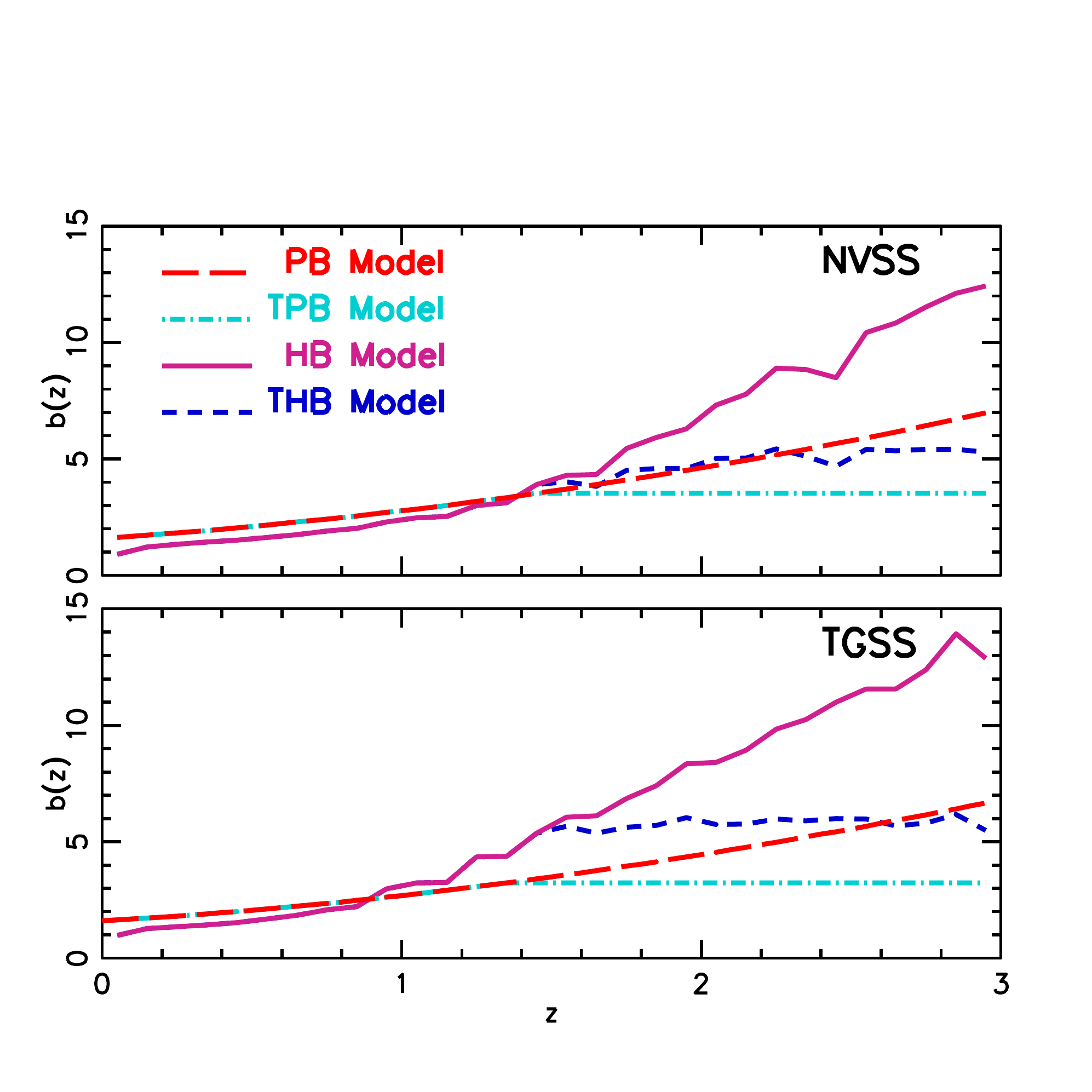}
\vspace{-1cm}
\caption{Effective bias function (Eq.~\ref{eq:beff}) for all the models listed in Table \ref{tab:model_params}.
The different bias models have different line styles, as indicated in the label.
Top panel: Bias function of  the {\it Reference} NVSS catalog.
Bottom panel: Bias function of the {\it Reference} TGSS catalog.
}
\label{fig:Bias_redshift_evolution_halo_masses}
\end{figure}

This parametric model relies on the physical model proposed by \cite{Nusser2015}.
Here we prefer to avoid using their parametric expression since the parameters were derived by constraining the correlation properties of the sources that they were investigating.
Instead, we follow the original \cite{Nusser2015} approach  and:
 {\it i)} assume that the radio activity is a strong function of the stellar mass,  {\it ii)} 
 adopt the expression provided by \cite{Nusser2015} to quantify the fraction of radio sources as a function of stellar mass and redshift, and {\it iii)} assume that the stellar mass is related to the halo mass as proposed by \cite{moster}.
The effective bias of the catalog can then be obtained by integrating equation 5 of \cite{Nusser2015}.
To do this we use the \cite{SMTbiasmodel}
halo bias model and the halo mass function of \cite{jenkins}. 
In this framework the difference between the TGSS and NVSS bias is determined by
the choice of the minimum halo mass that can host a radio source, which sets the lower limit of the integration.
For the NVSS case we adopt $1.4 \times 10^{11} M_\odot$, as in \cite{Nusser2015}, whereas for 
TGSS, which contains brighter objects, we use $10^{12} M_\odot$. 
However, as we have verified, this bias model is not very sensitive to the choice of this minimum mass.
The effective halo bias of the PB model  is represented by the red, long-dashed curves in Fig.~ \ref{fig:Bias_redshift_evolution_halo_masses}.

\item[•] 
\noindent \textit{Truncated Parametric Bias  model [TPB]}. It is the same as the PB model but, like in the THB case, we assume no bias evolution beyond $z=1.5$ The corresponding bias function is shown as the light blue, dot-dashed curve in 
Fig.~\ref{fig:Bias_redshift_evolution_halo_masses}.

\end{itemize}


\section{Angular power spectrum models versus data: $\chi^2$ comparison}
\label{sec:chisq analysis}

\begin{figure}
\vspace{-1.5cm}
\includegraphics[width=\columnwidth]{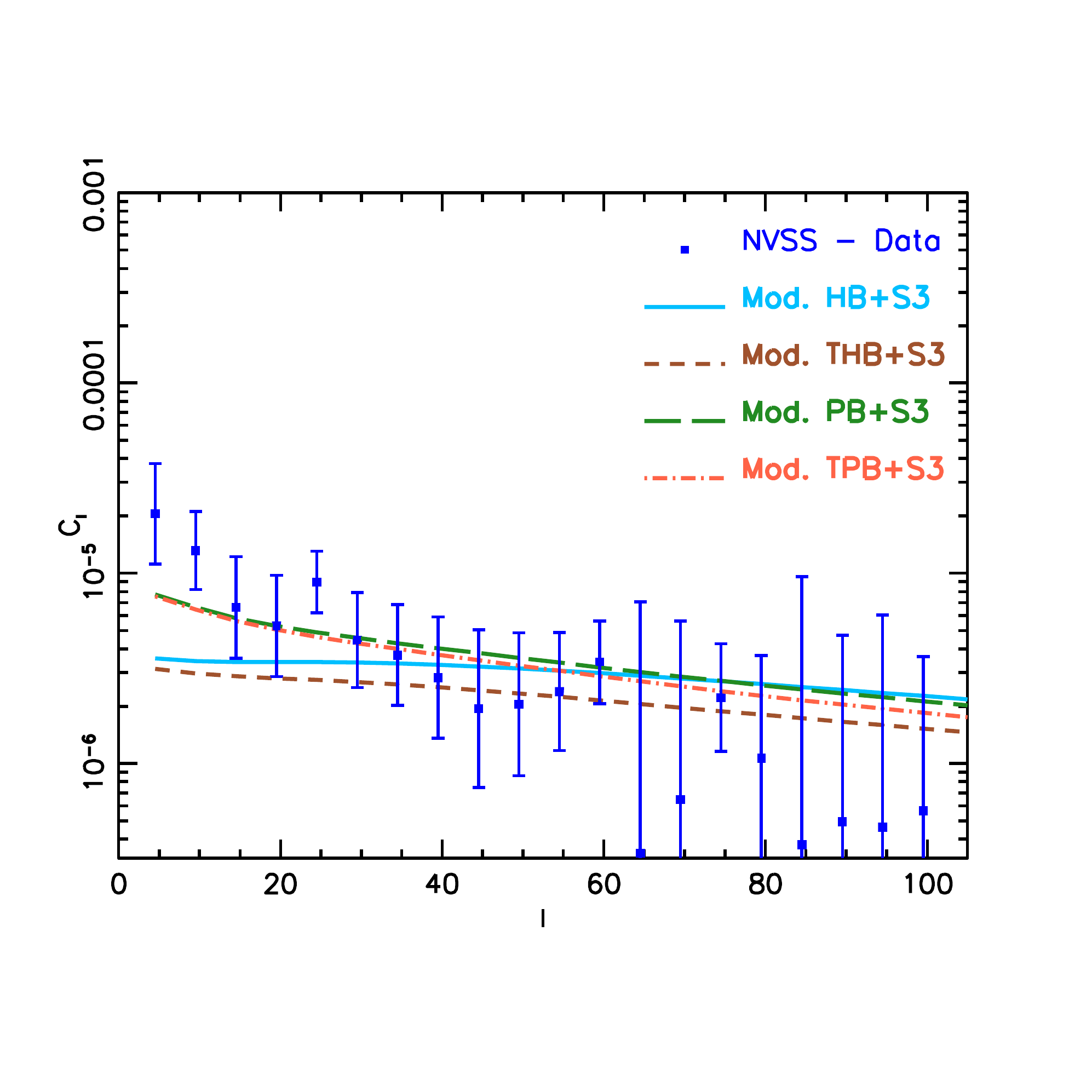}
\vspace{-1.6cm}
\caption{Measured NVSS APS (blue squares from Fig.~\ref{fig:Angular_spectra_TGSSvsNVSS}) vs. model predictions. The different models are listed in Table~\ref{tab:model_params} and described in the text, and represented with different linestyles, as indicated in the plot.}
    \label{fig:NVSS_modvsdat}
\end{figure}

\begin{figure}
\vspace{-1.5cm}
\includegraphics[width=\columnwidth]{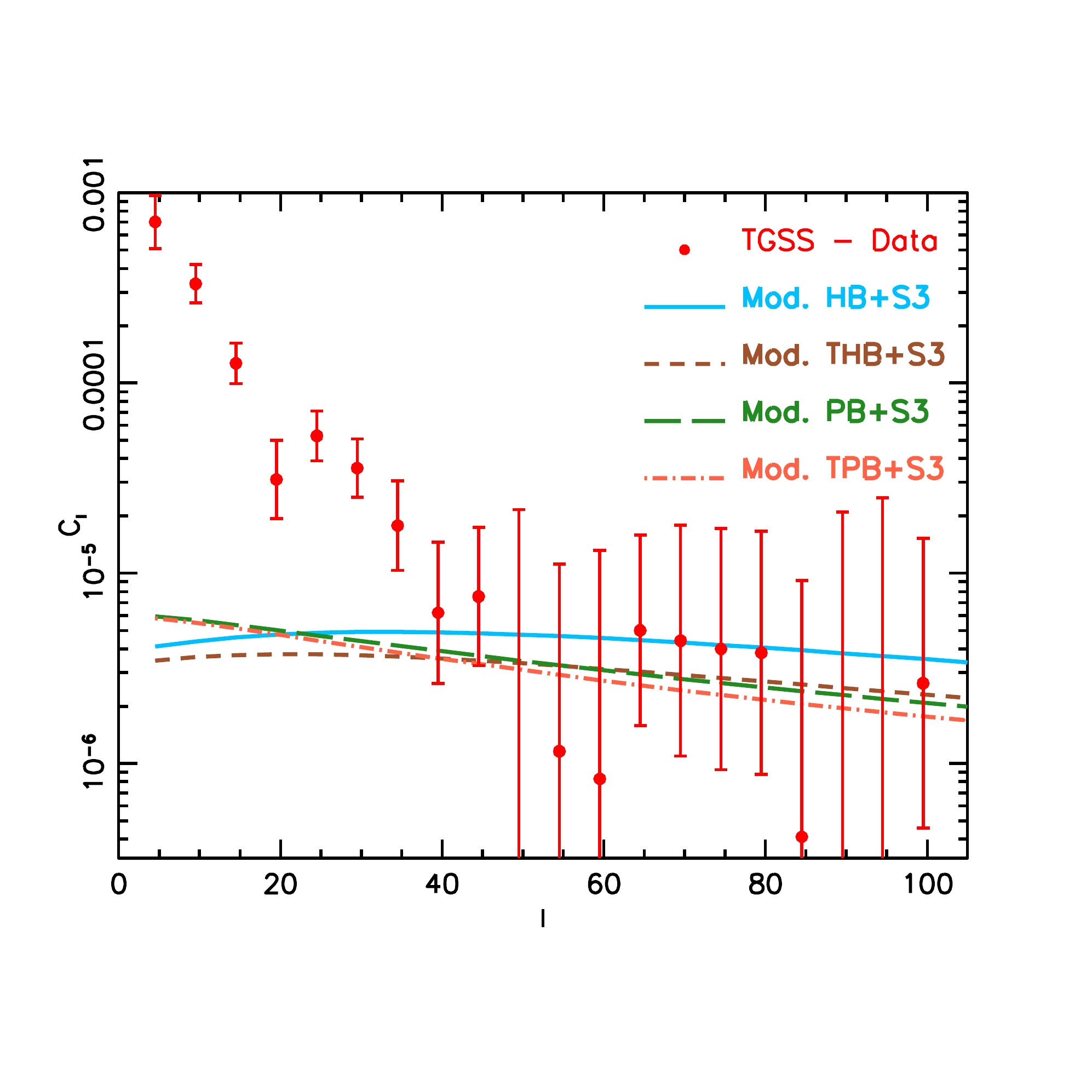}
\vspace{-1.6cm}
\caption{As in Fig.~\ref{fig:NVSS_modvsdat} but for the TGSS sample. The measured APS (red dots) is compared to model predictions (continuous curves with different linestyles). }
    \label{fig:TGSS_modvsdat}
\end{figure}

In Figures \ref{fig:NVSS_modvsdat} and \ref{fig:TGSS_modvsdat} we compare the measured NVSS and TGSS angular power spectra shown earlier in Fig. \ref{fig:Angular_spectra_TGSSvsNVSS}, with the APS models described in Sect. \ref{sec:modeling}.

\begin{table}
\center
\begin{tabular}{|c|c|c|c|} \hline \hline
Sample &  $ b(z)$ &$\chi^2_{30}/{\rm d.o.f.}$  $(Q=P(>\chi^2))$& $\chi^2_{\rm TOT}/{\rm d.o.f.}$ \\ \hline
\multirow{4}{*}{NVSS} & HB & $1.34$ (0.25) & $1.83$ \\
 & THB & $1.64$ (0.16)  & $1.21$  \\
 & PB & $0.61$ (0.65)  & $1.62$ \\
 & TPB & $0.66$ (0.62) & $1.30$  \\ \hline
\multirow{4}{*}{TGSS} & HB & $9.40$ ($4.5\times10^{-7}$) & $3.18$  \\
 & THB & $9.62$ ($2.9\times10^{-7}$) & $3.18$  \\
 & PB & $9.36$ ($4.7\times10^{-7}$) & $3.09$  \\
 & TPB & $9.42$ ($4.3\times10^{-7}$)  & $3.10$  \\
    \hline \hline
\end{tabular}\caption{Angular power spectrum model parameters used in the $\chi^2$ analysis and results.
Col. 1: Type of catalog. 
Col. 2: Bias model (see text for the meaning of the acronyms). Col. 3: reduced $\chi^2$ value obtained when considering the multipole range 
$\ell=[2,30]$ and the probability $Q=P(>\chi^2)$. Col. 4: reduced $\chi^2$ value obtained when considering the full multipole range 
$\ell=[2,100]$. In all the cases the redshift distribution $N(z)$ is based on the $S^3$ simulations as detailed in the text.}
\label{tab:model_params}
\end{table}

Already the visual inspection reveals that none of the APS models succeed in reproducing the 
angular power of TGSS sources at $\ell \leq 30$. The magnitude of the mismatch is remarkable indeed.
To quantify the discrepancy we have computed the reduced $\chi^2$ in two intervals: 
$\ell=[2,30]$ ($\chi^2_{30}$) to focus on the range in which the mismatch is larger and $\ell=[2,100]$ corresponding to the full multipole range considered in our analysis ($\chi^2_{\rm TOT}$). 
The $\chi^{2}$ was evaluated as follows (e.g. \citealt{dodelson}):
\begin{equation}
\chi^{2} = \sum_{\ell_1,\ell_2} (C_{\ell_1} - C^M_{\ell_1}) \mathcal{C}^{-1}_{\ell_1,\ell_2} (C_{\ell_2} - C^M_{\ell_2})
\end{equation}
where $C_{\ell}$ corresponds to the measured APS and $C_{\ell}^M$ to the APS model. We assume 
that the covariance matrix is diagonal, that is, $\mathcal{C}_{\ell_1,\ell_2} =\sigma_{C_{\ell_1}} \delta_{\ell_1,\ell_2}$, where $\sigma_{C_{\ell}}$ is the Gaussian error in Eq.~(\ref{errorbars}). 
The sum runs over all $\Delta_{\ell}$ bins from $\ell=2$ to either $\ell=30$ ($\chi^2_{30}$) or $\ell=100$ ($\chi^2_{\rm TOT}$).
The number of degrees of freedom $N_{\rm d.o.f.}$ is set equal to the number of $\Delta_{\ell}$ bins.
The values of the reduced $\chi^2$ are listed in Table \ref{tab:model_params} together with, for $\chi^2_{30}$ only, the probability $Q=P(>\chi^2)$.

We stress that here we are using the $\chi^2$ statistics to quantify the goodness of the fit, assuming no free parameters in the model APS.
The mismatch between prediction and measurement is so spectacular and the corresponding 
$\chi^2$ value is so large  that it is not
worth performing a more rigorous maximum likelihood analysis that accounts for error covariance
which, as we have argued, is expected to be small.
This result clearly shows that none of the physically motivated APS models built within the 
$\Lambda$CDM framework can account for the excess TGSS power on large scales, also when one 
takes into account theoretical uncertainties, quantified by the scatter in model predictions.

The only possibility to match the measured large-scale power would be to advocate a population of relatively local and highly biased radio sources that, however, is  neither 
supported by direct observational evidence nor by the results of the NVSS 
clustering analyses, which instead show that theoretical predictions 
match the measured APS,  as visible in 
Fig.~\ref{fig:NVSS_modvsdat}. The value of the reduced $\chi^2$ for NVSS in Table \ref{tab:model_params} is close to unity for all models explored and quantifies
the agreement in all the cases.
 
It is interesting to look at the differences between the APS models. At low redshifts the effective bias
of the PB and TPB models is larger than that of the HB and THB ones. This, and the fact that 
the angular power on large scales is largely generated locally (see e.g., Fig. 7 in \citep{Nusser2015}) , explains why the APS predicted by the 
PB models is larger than that predicted by the HB models at low multipoles, and why the former provide a better fit to the 
NVSS data. 
Also, truncating the bias evolution at $z=1.5$ has very little impact on our results since distant objects,
even if highly biased, are quite sparse and provide a shot-noise-like signal rather than producing coherent power on large angular scales.

It is worth pointing out that in this analysis we are considering the power within relatively large 
$\ell$ bins. Therefore, our result has no implication on the NVSS and TGSS dipole whose anomaly has been 
analyzed in a number of previous works. In this respect, all we can infer is that if indeed the 
NVSS dipole is anomalously large, then our analysis implies that the TGSS dipole is even larger, in qualitative agreement with the conclusions of the \cite{Bengaly2018} analysis.

\section{Discussion and Conclusions}
\label{sec:conclusion}

In this work we have analyzed the angular clustering properties of the radio sources in the First Alternative Data Release of the TGSS survey. Our analysis was performed in the harmonic space, to minimize error covariance and to facilitate the comparison with theoretical predictions, and focused on relatively large angular scales.
This choice was motivated by the results of recent clustering analyses that revealed a large clustering signal (compared to that of the NVSS sources) at angular separations larger than $\Delta \theta \simeq 0.1^{\circ}$ \citep{Rana2018} and an anomalously large dipole amplitude, in clear tension with $\Lambda$CDM 
expectations \citep{Bengaly2018}. Our aim was to investigate the behavior of the TGSS angular power spectrum 
at multipoles $\ell>1$ and compare it with theoretical predictions, taking into account known observational and theoretical uncertainties.
The clustering analysis of the TGSS sample was repeated on the NVSS catalog and on a sample of TGSS objects 
with a NVSS counterpart. The rationale behind this choice was to compare our results with those 
of a well-studied sample that contains most of the TGSS sources distributed over a similar 
sky area. 

The main results of our analysis are as follows.

\begin{itemize}
\item[•] The vast majority of TGSS sources have a counterpart in the NVSS catalog (about $94 \%$ when we consider 
our {\it Reference } samples) and are characterized by a spectral index Gaussian distributed around 
the value  $ \alpha_{\nu} \simeq -0.77$, similar to that of the NVSS sources and suggesting that the two catalogs contain similar classes of radio sources.

\item[•] The redshift distribution of TGSS sources extends well beyond $z=0.1$, that is, the typical scale 
probed by galactic counterparts with measured redshifts \citep{Rana2018}. We proved this point by cross-matching 
TGSS sources with optically identified QSOs in the SDSS-DR14 catalog. The fraction of cross-matched objects is 
small ($\sim 1.5$ \%) but sufficient to show that the distribution of TGSS sources extends beyond  $z=3$,
like the NVSS sources \citep{Nusser2015}.

\item[•] The angular two-point correlation function of TGSS sources exhibits a double power-law behavior,
qualitatively similar to that of the NVSS sources. Although not surprising, this result was not 
discussed by \cite{Rana2018} since they focused on angular scales 
larger than $0.1^{\circ}$. 
In that range the amplitude of the TGSS ACF is larger than that of the NVSS.
At small angles the behavior of the ACF is determined by the presence of radio sources with multiple 
components. We analyzed the behavior of the ACF on these small scales to quantify the clustering signal produced by 
multiple components and subtracted it from the measured angular power spectrum.

\item[•] The angular spectrum of TGSS sources has significantly more power than that of the NVSS in the multipole range $2 \leq \ell \leq 30$. Beyond $\ell=30$ the two spectra agree with each other within the errors.
This mismatch is also seen when the TGSS$\times$NVSS cross matched catalog is considered instead of the TGSS one. 

\end{itemize}

To check the robustness of this result to the known 
observational systematic errors we 
considered different TGSS samples obtained by varying the lower and upper flux selection thresholds and by using 
different geometry masks that exclude progressively larger regions of the sky near the Galactic plane.
The measured APS is remarkably robust to these changes and the TGSS versus NVSS power mismatch remains significant
even when going beyond the completeness limit of the TGSS catalog.
We did not explore the impact of errors in the flux calibration since these were found by \cite{Bengaly2018} to be small with respect to the magnitude of the mismatch.

Altogether these results excluded the hypothesis that the observed power mismatch could be attributed to known systematic errors related to the treatment of the data or to the observational strategy, and opened up the possibility that it may reflect 
genuine differences in the clustering properties of radio sources in the two catalogs.

To investigate this possibility we performed an absolute rather than a relative comparison 
between the measured TGSS angular spectrum and the one predicted in the framework of the $\Lambda$CDM model.
In doing this we took special care in modeling all the physical effects that contribute to the 
clustering signal and in propagating model uncertainties.
Among the physical effects, the ones that contribute the most to the large-scale clustering amplitude are
the redshift space distortions, which can boost the correlation signal by $\sim 3 \%$, and magnification lensing, which reduces the amplitude by $3-9 \%$, depending on the multipole considered. These effects were generally ignored in previous analyses. Although not 
negligible, their amplitude is far too small to explain the anomalous TGSS  power.
Finally, we find that the use of the Limber approximation, which has been adopted in many of the previous APS analyses, would spuriously enhance the predicted APS amplitude by $7-15 \%$, again depending on the multipole considered and being largest at $\ell <5$. 

The physical effects described above are well known and their contribution can be modeled with 
small errors. The largest uncertainties in modeling the TGSS spectrum are related to the composition of the catalog, the redshift distribution of its sources and, most of all, their bias.
To model the composition of the catalog and the redshift distribution of each source type we used the 
SKA Simulated Skies tool and found that our {\it Reference} TGSS catalog is mainly composed of FRII and FRI sources. Fainter radio objects like SFGs and GPS are comparatively fewer but very local, and therefore they represent a sizable fraction of the TGSS population at $z<0.1$. 
These objects are characterized by different redshift distributions and trace the underlying mass distribution with different biases. 

The biases of these sources and their evolution is the single most uncertain ingredient of our APS model.
To account for these uncertainties we considered four bias models. All of them assume a linear, deterministic biasing process and were conceived in the widely accepted framework of the "halo bias" model.
All of them are physically plausible, as they were designed to match the observed radio luminosity 
functions and number-count statistics. They differ from each other in the evolution of the bias beyond $z=1.5$ and in the 
relation between the radio sources and the mass of the host halo.

With all the ingredients and hypotheses previously described we generated four models for the APS of TGSS sources, and none of them are able to match the observed power at low multipoles. In fact the tension 
between models and data at $\ell \leq 30$ is so large that it makes a sophisticated error analysis unnecessary.
Our simple $\chi^2$ estimate is sufficient to reveal that the observed TGSS angular power spectrum cannot be
generated within the framework of a $\Lambda$CDM model: none of the physical effects described
are large enough to generate such a signal and none of the hypotheses on the nature, distribution and bias 
of TGSS radio sources can be stretched enough to simultaneously satisfy the luminosity function and the clustering properties of these sources.

It is remarkable that, instead, our models match the angular spectrum of NVSS galaxies in the same range
$2 \leq \ell \leq 30$ once the observational errors and theoretical uncertainties are taken into account.
This result confirms that our APS models are indeed physically viable.

We are left with the uncomfortable evidence of an excess large-scale clustering in the angular distribution
of the TGSS ADR1 sources. The excess is seen both in the comparison with similar analyses carried out on 
the NVSS dataset which shares many similarities with TGSS, and in the comparison with theoretical predictions.
In Section \ref{sec:APS} and, more extensively, in the Appendix we 
searched for possible observational effects that may generate a spurious clustering signal large enough to 
explain the tension detected by our study but failed to identify an obvious candidate.

There is an obvious continuity between our results and those of \cite{Bengaly2018} who detected an 
anomalous large amplitude in the $\ell=1$ dipole moment of the TGSS angular spectrum. For this reason 
we agree with their conclusion that the observed mismatch indicates the presence of unidentified 
systematics in the data not captured by the ones that we have explicitly searched for in this study.
It may be that this issue can only be clarified with future TGSS data releases or thanks to other forthcoming wide-angle radio surveys carried out at similar frequencies like the ongoing LOFAR Two-Metre Sky Survey (LOTSS; \citet{Shimwell2017,Shimwell2018}) and/or future, deeper releases of the GLEAM catalog \citep{HurleyWalker2017,White2018}.

\begin{acknowledgements}
We would like to thank Lee Whittaker for his help in modeling the magnification lensing effect.
We also thank Chris Blake, Prabhakar Tiwari, and the anonymous referee for their useful comments and suggestions.
EB is supported by MUIR PRIN 2015 ``Cosmology and Fundamental Physics: illuminating the Dark Universe with Euclid'' and Agenzia Spaziale Italiana agreement ASI/INAF/I/023/12/0. MB, IP and EB acknowledge support of the Ministry of Foreign Affairs and International Cooperation, Directorate General for the Country Promotion (Bilateral Grant Agreement ZA14GR02 - Mapping the Universe on the Pathway to SKA). IP and EB also acknowledge support from INAF under the SKA/CTA PRIN {\it FORECaST}. MB is supported by the Netherlands Organization for Scientific Research, NWO, through grant number 614.001.451 and by the Polish Ministry of Science and Higher Education through grant DIR/WK/2018/12. ABA acknowledges financial support from the Spanish Ministry of Economy and Competitiveness (MINECO) under the Severo Ochoa program SEV-2015-0548. 
\end{acknowledgements}



\bibliographystyle{mnras}
\bibliography{biblio} 


 
\begin{appendix}
\section{Observational effects potentially affecting the APS estimate}

To search for possible observational effects that may affect our APS measurement we performed a number of tests to detect potential, observationally driven sources of spurious clustering signals,
in addition to those presented in Sections \ref{sec:fcuts} and \ref{sec:gcuts}.

\begin{figure}
\includegraphics[width=\columnwidth]{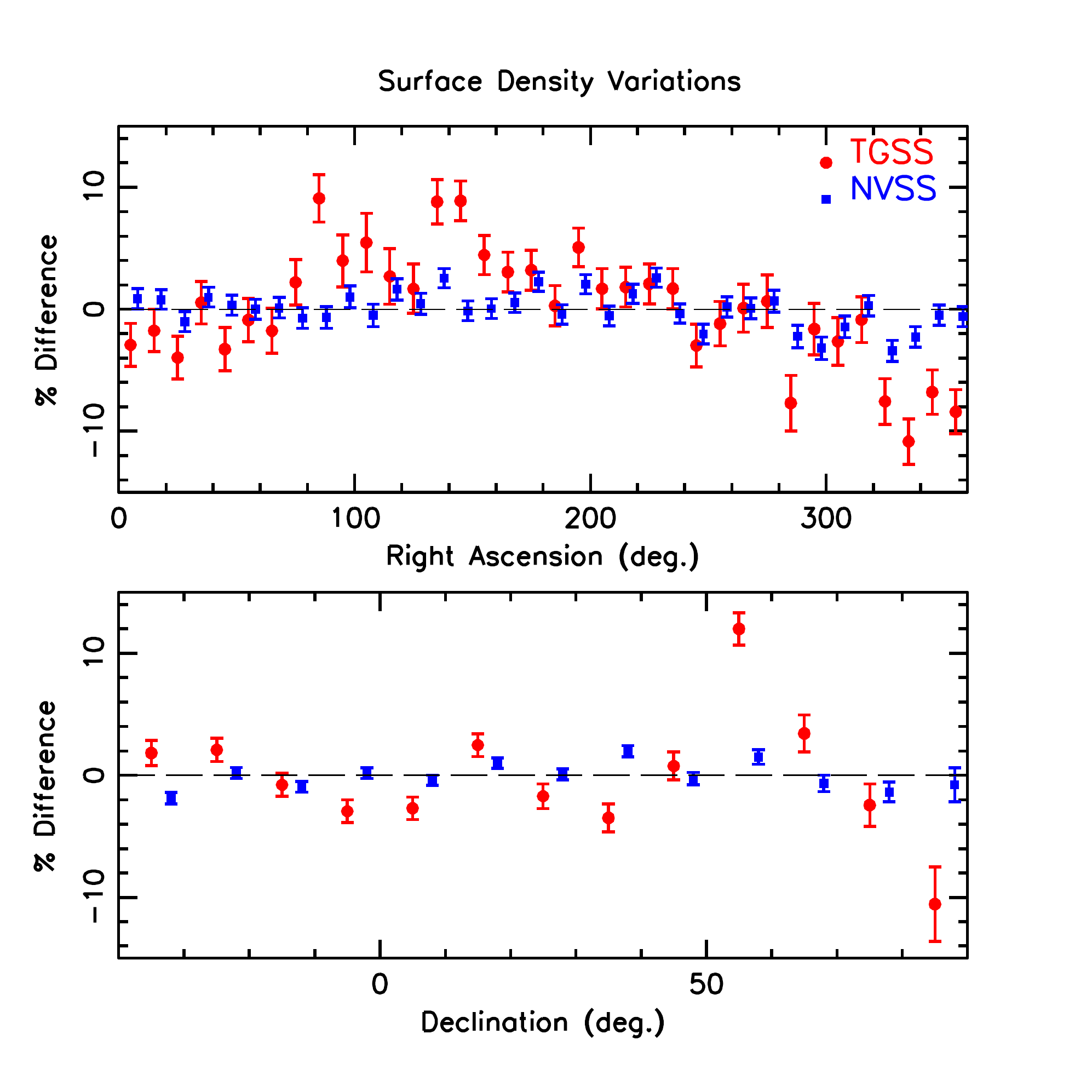}
\caption{Top panel: percentage variations in the surface density of TGSS (red dots) 
and NVSS (blue squares) sources as a function of RA. 
Bottom panel: percentage variations as a function of DEC. Error bars 
represent Poisson noise in the corresponding stripe.}
    \label{fig:surface}
\end{figure}

In the first of such tests we repeated the same analysis as \cite{BW02b} and searched for variations
in the surface density of TGSS objects as a function of right ascension (RA) and declination (DEC). The mean surface density has been computed over the same area as the {\it Reference} catalog and compared with the same quantity estimated over stripes of constant RA and DEC. In Figure \ref{fig:surface} we show the corresponding percentage surface density variations of the TGSS objects (red dots) and of
NVSS sources (blue squares) in their {\it Reference} catalogs. The error bars are Poisson $\sqrt{N}$ of the objects in each stripe.
Surface density variations in the TGSS survey are significantly larger than in the NVSS catalog;
they show a sinusoidal dependence  on  DEC and, more prominently, on RA.
In the upper panel the surface density has a minimum ($\sim -8$ \%) in correspondence of the RA interval $[300^{\circ},360^{\circ}]$ 
and a maximum ($\sim +8$ \%) in the RA interval $[90^{\circ},140^{\circ}]$.
In the bottom plot the sinusoidal trend is less evident, with fluctuations of about 2\% except that at DEC $>+60$\% where variations larger than 10 \% are detected. 

\begin{figure}
\includegraphics[width=\columnwidth]{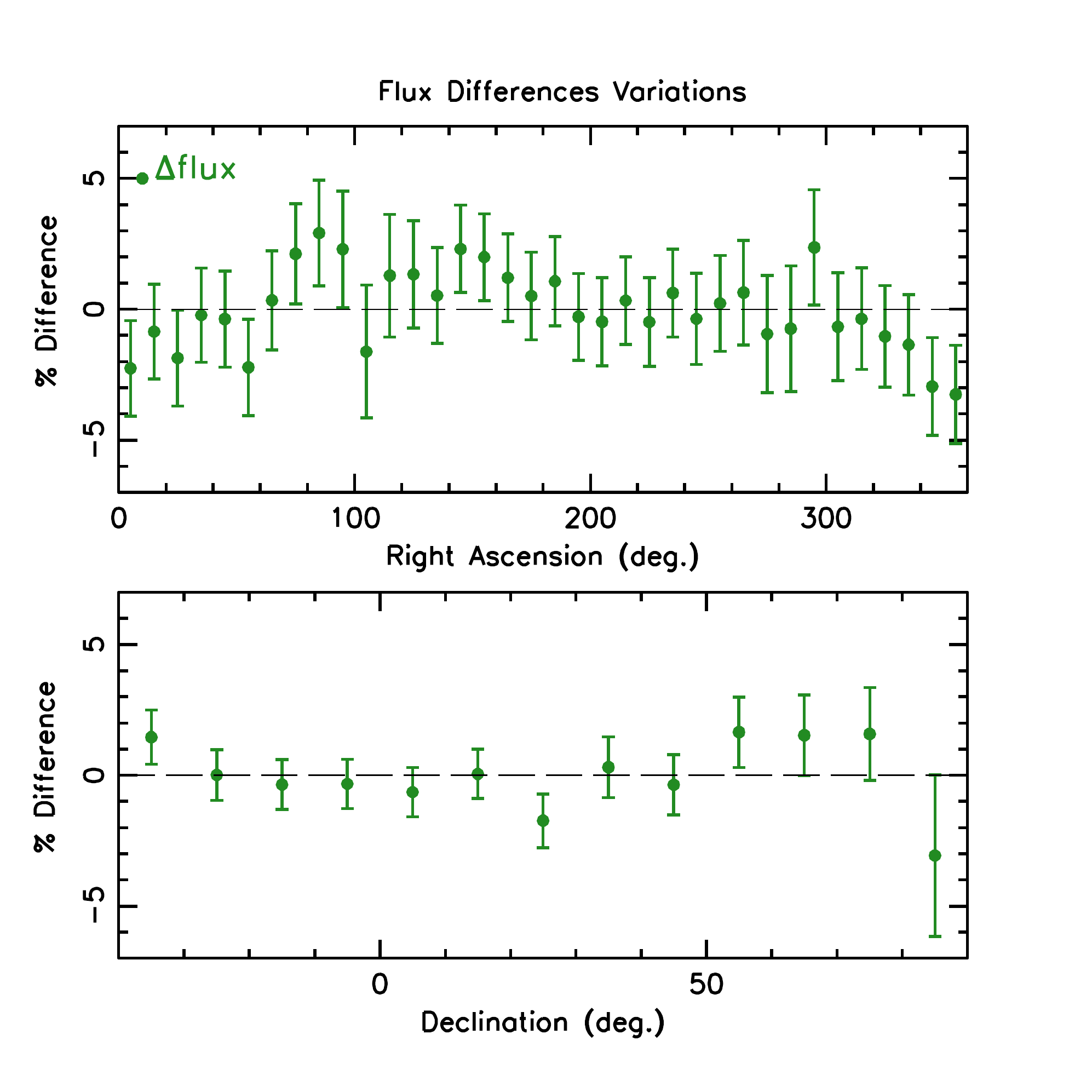}
\caption{Percentage variations in the difference of TGSS-NVSS flux for 
all sources in the TGSS$\times$NVSS catalog. The flux differences are computed 
as a function of the RA (top panel) and DEC (bottom panel). Error bars show the 
{\it RMS} scatter of the quantity estimated in the pixels of each RA and DEC stripe.}
    \label{fig:flux}
\end{figure}

What is the origin of these variations? 
Surface density fluctuations may reflect systematic errors in the flux measurements. To test this hypothesis we have computed the difference between the TGSS and NVSS flux and searched for correlation with the RA and DEC of the sources.
The results are shown in  Figure \ref{fig:flux}. 
A sinusoidal behavior similar, though less significant, to that of the surface density fluctuation, is seen as a function of RA (upper panel). \cite{Intema2017} found that systematic pointing offsets in the TGSS survey generate variations in the measured flux whose amplitude depends on the RA. This effect has been modeled and corrected for by the authors. However, the fact that we observe a residual dependence on the RA of the sources seems to indicate that the correction may not be exact. 

Flux variations as a function of declination are also present, but weaker, and less obviously correlated to surface density fluctuations. As discussed by \cite{Intema2017}, such systematic errors can arise from the fact that the synthesized beam size depends on the declination of the source. This effect was also corrected for in the TGSS catalog \citep{Intema2017}. However, just like for the RA case, our results seems to indicate that some systematic effect is still present in the data. 
 
Overall, these results provide a weak evidence that surface density fluctuations are determined by or related to systematic errors in the flux measurements.

Are these surface density fluctuations responsible for the anomalous large-scale power that we have detected in the APS of TGSS sources? 
As \cite{BW02b} pointed out, a 10 \% surface density shift generates an offset $\Delta w \sim 0.01$ in the angular correlation on scales less than that on which the surface density varies. Therefore, the $\ga 10\%$ fluctuations seen at 
$\delta>+60^{\circ}$  are in principle consistent with the observed mismatch in the angular correlation functions of NVSS and TGSS
beyond $\theta=0.1^{\circ}$ shown in Fig.~\ref{fig:TGSS_autocorrelation_function_compare_samples}.

\begin{figure}
\vspace{-1.5cm}
\includegraphics[width=\columnwidth]{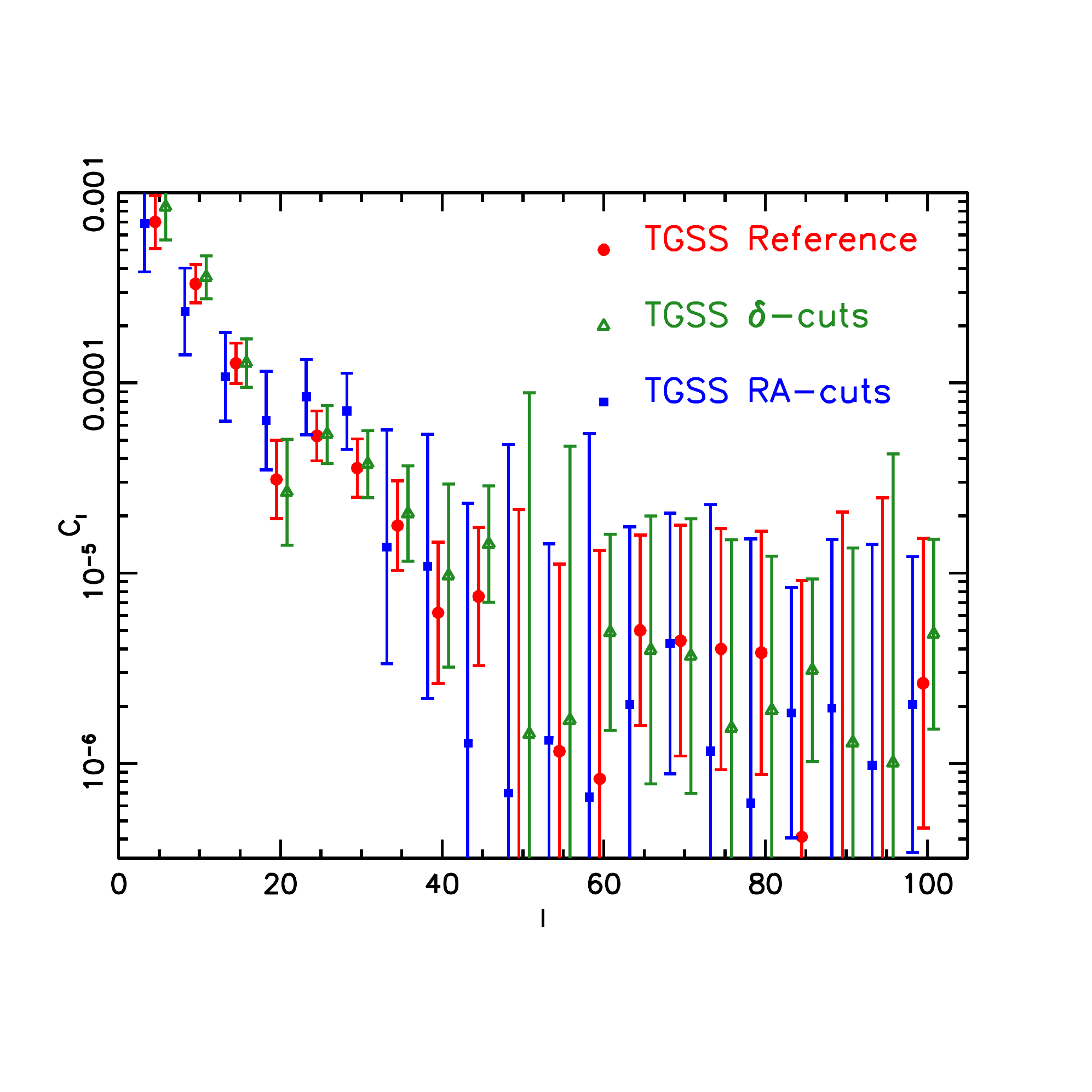}
\vspace{-1.2cm}
\caption{Angular power spectrum of the {\it Reference} sample (red dots) compared to 
that of a subsample of TGSS sources in the region $-20^{\circ}<\delta<+60^{\circ}$ (green triangles) and of a subsample of TGSS sources in the RA-range $[200^{\circ},280^{\circ}]$ (blue squares). Error bars represent Gaussian errors. A small horizontal offset has been applied to the green triangles and blue squares to avoid overcrowding.
}
\label{fig:delta}
\end{figure}

To quantify the impact of the observed RA and DEC surface density fluctuations on the measured angular spectrum we extracted TGSS subsamples by selecting regions in which surface density fluctuations are small
and measured the corresponding  APS.
First of all we considered the declination dependence. In this case the TGSS subsample was obtained
by excluding all sources with DEC $>+60^{\circ}$, to exclude regions with large surface density, and 
with DEC $<-20^{\circ}$, to exclude a region with moderate surface-density variations, which, however, could provide the necessary leverage to mimic clustering on very large angular scales.
Figure \ref{fig:delta} shows that the angular power spectrum of this TGSS sub-sample (green triangles) agrees with that of the {\it Reference} sample (red dots), despite their different angular masks. 
Second of all, we focused on the RA-dependence. We performed an even more aggressive cut and considered 
objects in the RA range $[200^{\circ},280^{\circ}]$ where the surface-density fluctuations are small.
The APS of this TGSS subsample is represented by blue squares in Fig. \ref{fig:delta}. Also in this case,
the result agrees with the {\it Reference} sample.
Together, these results indicate that the DEC- and RA-dependent surface-density variations cannot be responsible for the excess large-scale power of the TGSS sample.

\begin{figure}
\vspace{-1.5cm}
\includegraphics[width=\columnwidth]{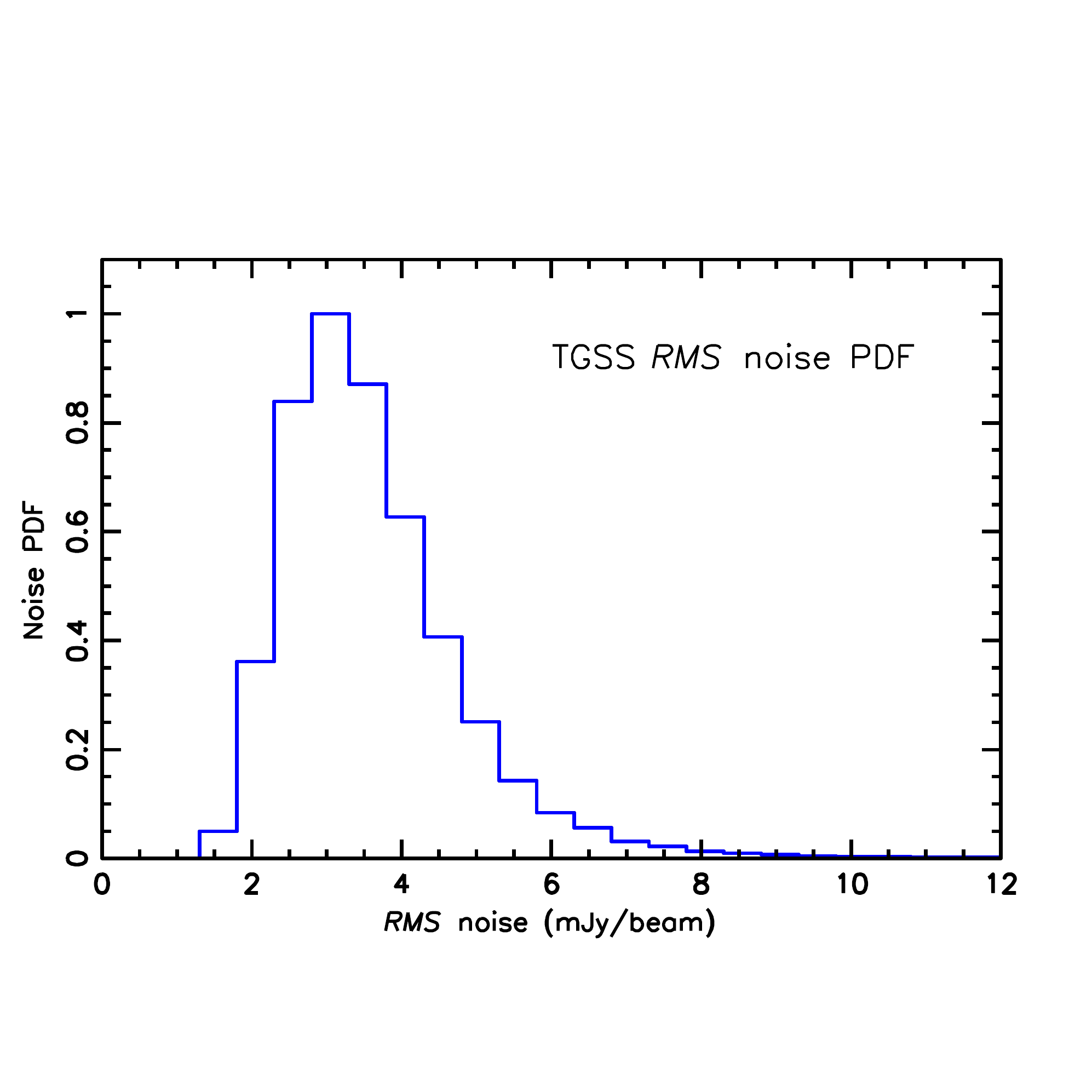}
\vspace{-1.6cm}
\caption{Distribution of the {\it RMS} noise in the TGSS {\it Reference} catalog.}
    \label{fig:histo2}
\end{figure}

Subsequently, we analyzed the properties of {\it RMS}  noise in the TGSS catalog. 
The noise was estimated over an area of 40$\times$40 
pixels centered at the position of  each TGSS source.
According to \cite{Intema2017} this noise is mainly contributed  by the thermal receiver noise and by the image noise  due to sparse UV coverage. Therefore it should not correlate with
the clustering and other properties of the TGSS sources.
To investigate the validity of this hypothesis we performed several tests.
First of all we checked that no correlation exists between the noise and the flux of the sources in the {\it Reference } catalog.
Second, we computed the probability distribution of this {\it RMS} noise (shown in Fig.~\ref{fig:histo2})
and compared it to that of the noise measured in the mosaic images (Fig. 7 in \citealt{Intema2017}). We found that the two distributions have similar median values (3.77 vs. 3.5 mJy/beam) and shapes
(the fraction of points with noise smaller than 5 mJy/beam is 86.7 \% in our TGSS {\it Reference} sample and 80 \% in the mosaic images). 
We conclude that the properties of the noise at the location of the sources is statistically equivalent to that measured over the whole survey area.

\begin{figure}
\includegraphics[width=\columnwidth]{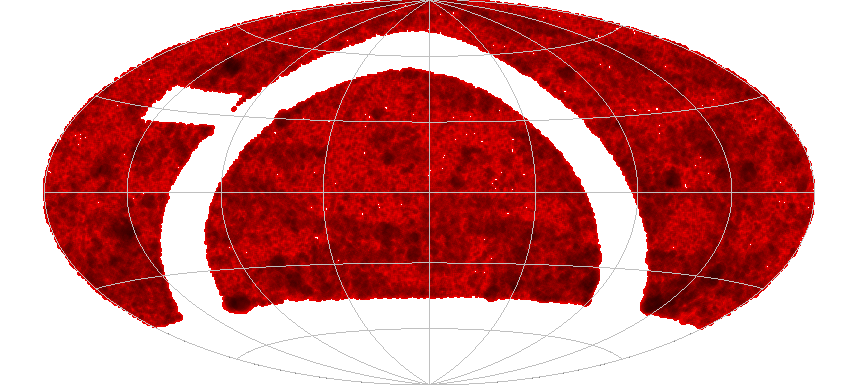}
\vspace{-0.4cm}
\caption{Aitoff projection of the {\it RMS} noise measured at the location of TGSS sources. The size of each datapoint is weighted by the noise amplitude. Darker spots indicate low-density regions of TGSS objects and/or low {\it RMS} noise.
}
    \label{fig:Noisemap}
\end{figure}

In the third set of tests we investigated the angular correlation properties of the TGSS noise. The sky distribution of this noise (measured at the source locations) 
is shown in Figure \ref{fig:Noisemap} in the same coordinate system as in Fig.~\ref{fig:TGSS_NVSS_maps}. 
Since in the map the size of each dot is proportional to the noise value,
dark (bright) areas may either represent regions of low (high) source density or of
low (high) noise.
Visual inspection reveals a significant degree of correlation on large angular scales.
To understand if this correlation simply reflects the correlation properties of the underlying TGSS sources or if, instead, it flags the presence of a spurious correlation between the {\it RMS} noise and the
clustering of the TGSS sources, we measured the cross angular power spectrum [XPS] between the noise and the source TGSS maps and compared it with the APS of the TGSS sources.
Since the noise is measured at the position of the TGSS sources, a match between the XPS and the APS of the 
TGSS sources would exclude a correlation between noise and TGSS clustering.

\begin{figure}
\includegraphics[width=\columnwidth]{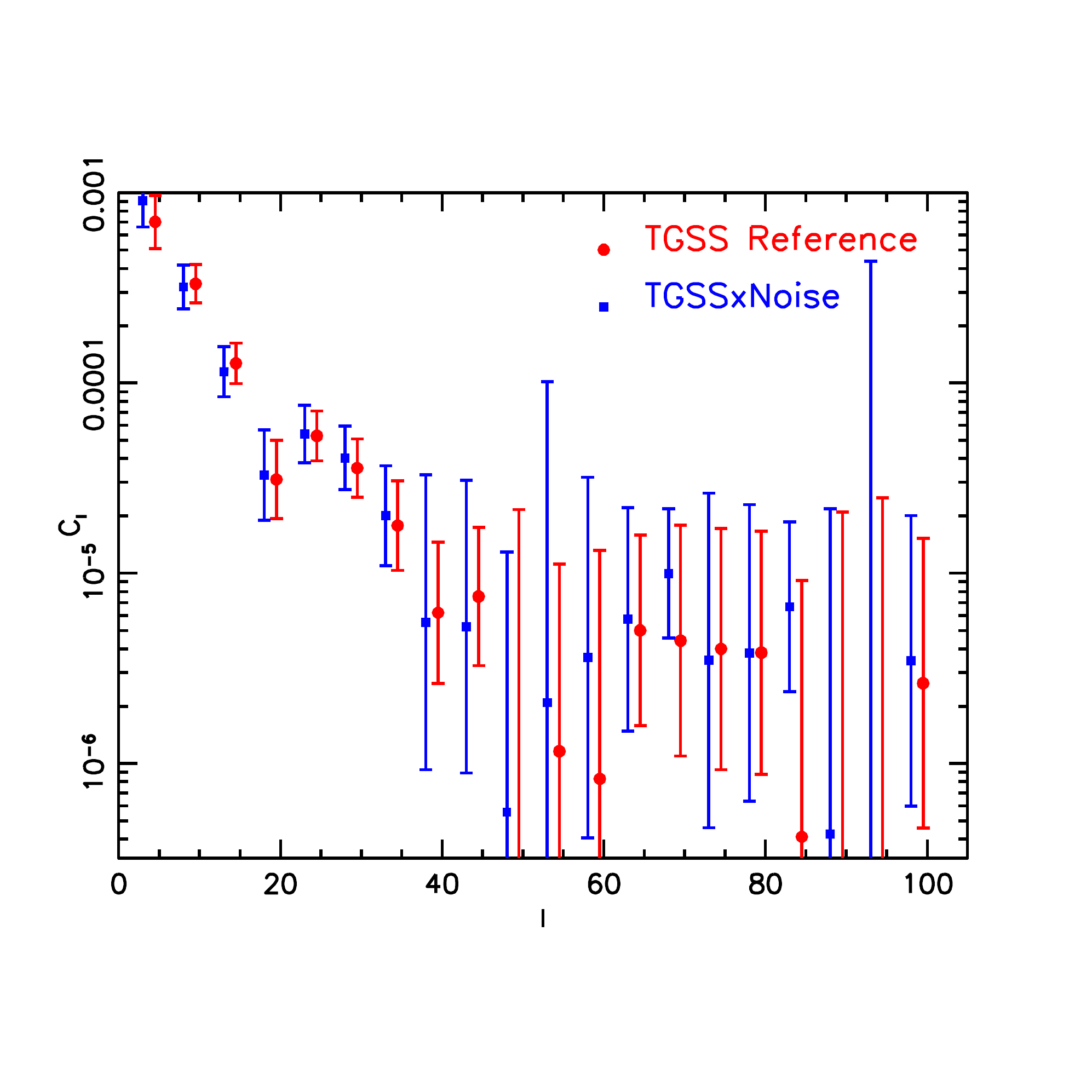}
\vspace{-1.6cm}
\caption{Angular power spectrum of the {\it Reference} sample (red dots) compared to 
the {\it RMS} noise-TGSS cross-spectrum (blue squares).
 Error bars represent Gaussian errors. A small horizontal offset has been applied to the blue squares to avoid overcrowding.
}
\label{fig:XPS}
\end{figure}

The result shown in Fig. \ref{fig:XPS}, indicates that the cross-spectrum (blue squares) is consistent with the  APS of the {\it Reference} TGSS sources (red dots). We also found that the APS of the noise (not shown) 
agrees with that of the TGSS sources. We therefore conclude that the excess TGSS large-scale power is not
induced by correlations with the {\it RMS} noise.

\begin{figure}
\includegraphics[width=\columnwidth]{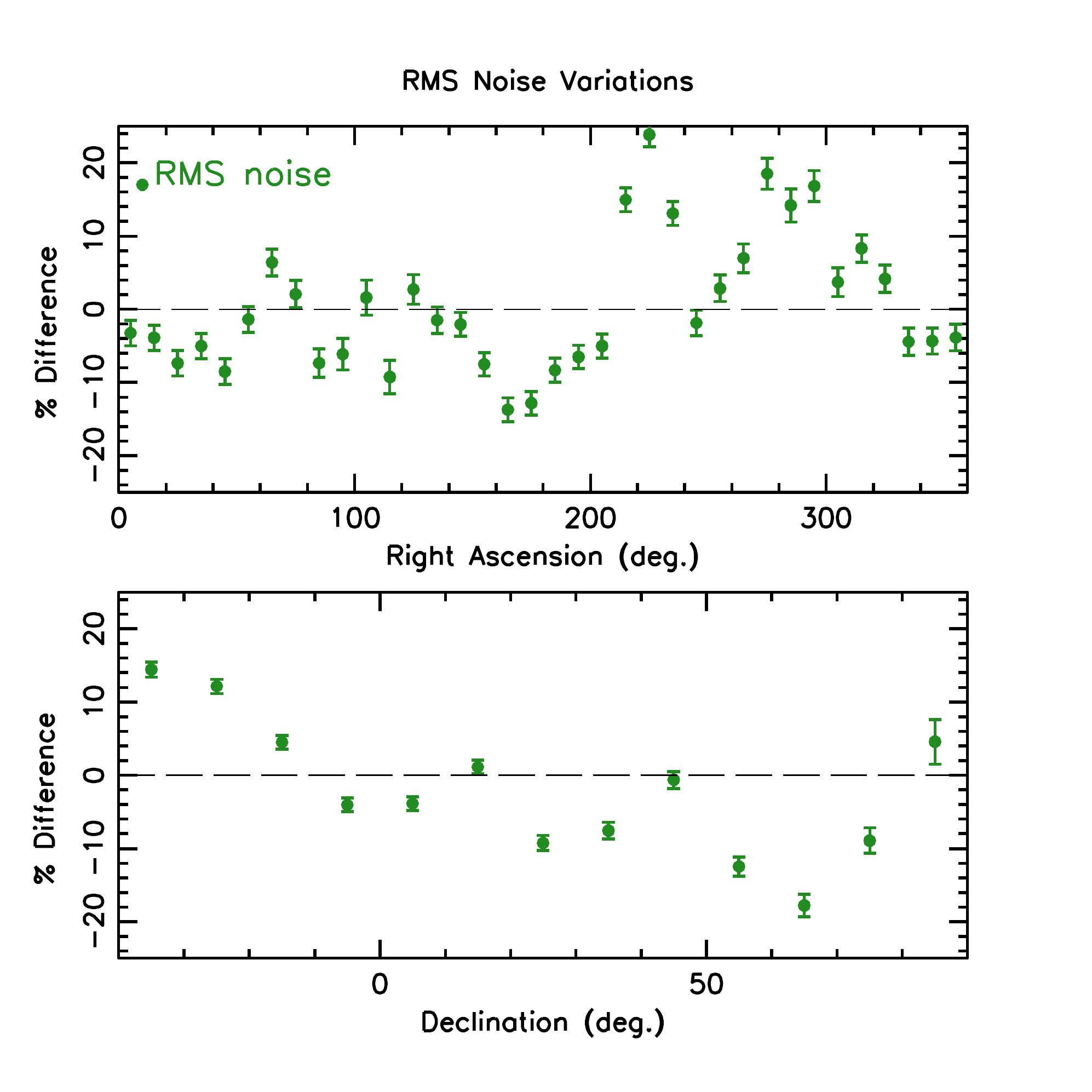}
\caption{Percentage variations in the RMS noise of TGSS sources as a function of RA (top panel) and DEC (bottom panel). Error bars represent Poisson noise in the corresponding stripe.}
    \label{fig:RMS}
\end{figure}

Finally, we repeated the same test as in Fig. \ref{fig:surface} and 
searched for noise dependence on the RA (top panel of figure \ref{fig:RMS})
and on the DEC (bottom panel).
We detected variations as large as 20 \% that, however, do not match those 
of the source surface density. 
These tests reveal that TGSS noise presents intrinsic angular correlation properties that, however,
do not seem to be obviously related to those of the TGSS sources. This is likely another confirmation of the fact that the flux threshold adopted to define the TGSS reference catalog ($S>200$ mJy) is conservative enough to avoid incompleteness effects associated with noise variations.

\label{sec:appA}
\end{appendix}


\end{document}